\documentclass[twoside]{article}
\usepackage{qic,epsfig,bbm,amsmath}

\renewcommand{\thefootnote}{\fnsymbol{footnote}}  %use symbolic footnote

\begin{document}
\setlength{\textheight}{8.0truein}    %FOR 2ND PAGE ONWARDS

\runninghead{Analysis of circuit imperfections in BosonSampling}
            {Anthony Leverrier, Ra\'ul Garc\'ia-Patr\'on}

\normalsize\textlineskip
\thispagestyle{empty}
\setcounter{page}{1}

%\copyrightheading{Vol.}{No.}{Year}{Page Nos.}
\copyrightheading{0}{0}{2003}{000--000}

\vspace*{0.88truein}

\alphfootnote

\fpage{1}

\centerline{\bf ANALYSIS OF CIRCUIT IMPERFECTIONS IN BOSONSAMPLING}

\vspace*{0.37truein}
\centerline{\footnotesize
%%%%%%%%%%%%%%%%%%%%%%%%%%%%%%%%%%%%
%put authors' name and address here
%%%%%%%%%%%%%%%%%%%%%%%%%%%%%%%%%%%%
ANTHONY LEVERRIER}
\vspace*{0.015truein}
\centerline{\footnotesize\it INRIA Rocquencourt, Domaine de Voluceau}
\baselineskip=10pt
\centerline{\footnotesize\it B.P. 105, 78153 Le Chesnay Cedex, France}
\vspace*{10pt}
\centerline{\footnotesize 
RAUL GARCIA-PATRON}
\vspace*{0.015truein}
\centerline{\footnotesize\it Quantum Information and Communication, Ecole Polytechnique de Bruxelles, CP 165, Université Libre de Bruxelles}
\baselineskip=10pt
\centerline{\footnotesize\it 1050 Bruxelles, Belgium}
\centerline{\footnotesize\it Max-Planck Institut fur Quantenoptik, Hans-Kopfermann Str. 1}
\baselineskip=10pt
\centerline{\footnotesize\it D-85748 Garching, Germany}
\vspace*{0.225truein}
\publisher{(received date)}{(revised date)}

\vspace*{0.21truein}

%% \abstracts{first paragraph}{second paragraph}{third paragraph}
%% If there is only one paragraph, just keep the second and third empty 
%% like the following one 
\abstracts{
\textsc{BosonSampling} is a problem where a quantum computer offers a provable speedup over classical computers.
Its main feature is that it can be solved with current linear optics technology, without the need for a full quantum computer.
In this work, we investigate whether an experimentally realistic BosonSampler can really solve \textsc{BosonSampling} without any fault-tolerance mechanism.
More precisely, we study how the unavoidable errors linked to an imperfect calibration of the optical elements affect the final result of the computation. 
We show that the fidelity of each optical element must be at least $1 - O(1/n^2)$, where $n$ refers to the number of single photons in the scheme. Such a requirement seems to be achievable with state-of-the-art equipment.  
}{}{}

\vspace*{10pt}

\keywords{The contents of the keywords}
\vspace*{3pt}
\communicate{to be filled by the Editorial}

\vspace*{1pt}\textlineskip    %) USE THIS MEASUREMENT WHEN THERE IS
   %) A SECTION HEADING
%\vspace*{-0.5pt}
%\noindent
%%%%%%%%%%%%%%%%%%%%%%%%%%%%%%%%
%put the text of the paper here
%%%%%%%%%%%%%%%%%%%%%%%%%%%%%%%%

\section{Introduction}

While quantum computers are widely believed to provide speed-ups over classical computers in theory, it remains an outstanding experimental challenge to provide hard evidence for such a speed-up. Shor's factoring algorithm \cite{Sho97} would be a natural candidate algorithm, but suffers from two main caveats: first, one would need a full-blown quantum computer in order to factor integers large enough to beat any classical competition (provided one does not know the factors in advance \cite{SSV13}); second, it is not known whether factoring can be performed efficiently on a classical computer. Arguably, the first reason is already quite compelling: building a universal quantum computer turns out to be a very difficult task. 
A much sounder experimental proposal to establish \emph{quantum supremacy} \cite{Pre12} was recently introduced by Aaronson and Arkhipov as a problem called \textsc{BosonSampling}\footnote{We write \textsc{BosonSampling} when referring to the complexity problem, and ``Boson Sampling'' when referring to the optical experiment.} \cite{AA13}. In principle, this problem can be solved with a simple linear optics experiment by processing a bunch of single photons through a network of passive optical elements, namely beamsplitters and phase-shifters, and counting the number of photons in the output modes. This experimental simplicity certainly explains why the proposal has attracted a lot of attention recently \cite{BFR13,TDH13,SMH13,COR12}.

Boson Sampling appears as a simple generalization of the two-photon Hong-Ou-Mandel effect, well-known in quantum  optics \cite{HOM87}. More precisely, given an arbitrary $m$-mode passive optical network, the goal is to process an initial state consisting of single photons in $n$ of the modes and vacuum in the remaining modes, and to measure the distribution of photons in the output modes. 
Despite the absence of obvious applications, this task provides evidence that an allegedly rudimentary quantum computer can outperform its classical counterpart. Indeed, were an \emph{efficient} classical algorithm for \textsc{BosonSampling} to exist, the Polynomial Hierarchy \cite{Sto76} would collapse at the third level, something believed to be highly implausible in complexity theory. To be more precise, the existence of an efficient classical algorithm for \emph{exact} \textsc{BosonSampling} implies the collapse of the Polynomial Hierarchy, and it is conjectured that the existence of an efficient classical algorithm for approximate \textsc{BosonSampling} would have a similar implication (see Ref.~\cite{AA13} for a precise statement of the conjecture).

Let us now describe more precisely the setup of Boson Sampling and discuss some experimental perspectives.  The \textsc{BosonSampling} problem is characterized by two parameters: the number $n$ of single photons and the number $m$ of optical modes. Typically, one chooses $m$ to depend quadratically on $n$ since an application of the bosonic birthday paradox \cite{AK11} shows that in this regime, most of the output modes will either contain 0 or 1 photon\footnote{The regime where the number of modes is linear in the number of photons is also interesting but remains mostly unexplored, partly because it is more difficult to analyze.}.
An instance of the \textsc{BosonSampling} problem is described by the choice of a unitary $U^{\mathrm{BS}}$ acting on $\mathbbm{C}^m$. This unitary characterizes the $m$-mode interferometer that should be used in the experiment, and it is typically drawn according to the Haar measure on the unitary group $U(m)$. 
The output of the \textsc{BosonSampling} problem is a sample of a certain probability distribution depending on $U^\mathrm{BS}, n$ and $m$ which we describe now. Consider the $m$-mode optical state $|1\rangle^{\otimes n} \otimes |0\rangle^{\otimes (m-n)}$ containing a single photon in each of the first $n$ modes and vacuum in the remaining ones, and send it through an array of beamsplitters and phase-shifters implementing the optical unitary transformation corresponding to $U^\mathrm{BS}$. The probability distribution one should sample from corresponds to the distribution of the $n$ photons in the $m$ \emph{output} modes of the optical network.

\begin{figure}
\centering
  \includegraphics[width=.8\linewidth]{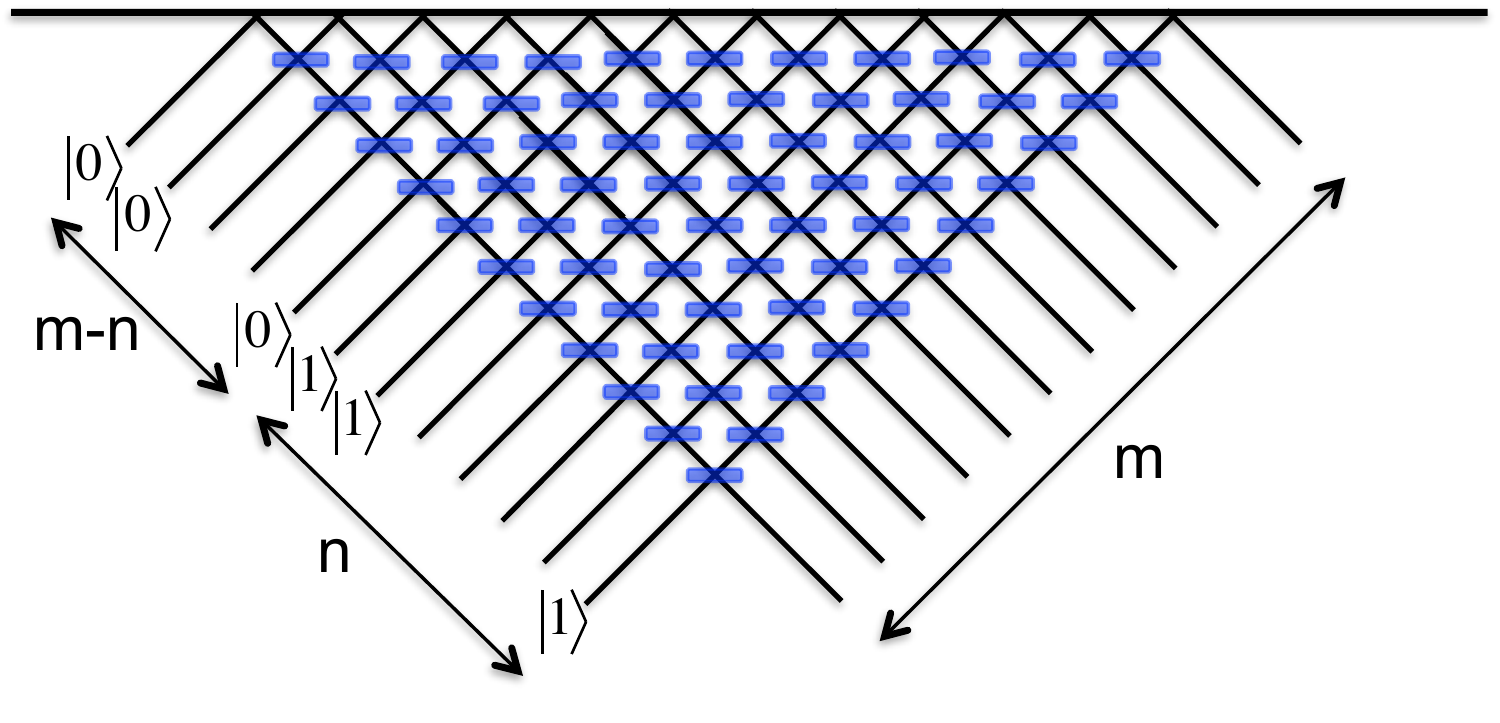}
  \label{fig:scheme1}
\fcaption{Implementation of the Boson Sampling experiment, following the scheme of Ref.~\cite{RZB94}. In general, one would need both phase-shifters and beamsplitters. Here, for simplicity, we consider beamsplitters with complex entries. In other words, phase-shifters are absorbed in the beamsplitters.}
\label{fig:scheme}
\end{figure}

From the description of the problem, it is clear how to solve \textsc{BosonSampling} with the help of a programmable array of beamsplitters, sources of single photons and photon counters (see Fig.~\ref{fig:scheme}): simply implement the optical network corresponding to $U^\mathrm{BS}$, send undistinguishable single photons in the first $n$ modes and count the photons in the output modes. 
We point out that if the annihilation operators of the $m$ input modes are given by the vector $\vec{a} = (a_1, \ldots, a_m)$, those of the output modes correspond to $U^\mathrm{BS} \vec{a}$.

On the other hand, sampling from this distribution with classical means only seems quite difficult. Indeed, the probability $\mathrm{Pr}[\vec{s}]$ of obtaining a given output sequence $\vec{s} = (s_1, \ldots, s_m)$ where the integer $s_i$ corresponds to the number of photons in the $i^\mathrm{th}$ output mode is given by
\begin{align}
\label{proba_permanent}
\mathrm{Pr}[\vec{s}] = \frac{|\mathrm{Per} (U_{\vec{s}}^\mathrm{BS})|^2}{s_1! \cdots s_m!},
\end{align}
where $\mathrm{Per}$ is the permanent and $U_{\vec{s}}^{\mathrm{BS}}$ is the $n \times n$ matrix obtained from $U^\mathrm{BS}$ by discarding all but the $n$ first columns of $U^\mathrm{BS}$ and then, for all $i \in \{1, \ldots, m\}$, taking $s_i$ copies of the $i^\mathrm{th}$ row of $U^\mathrm{BS}$ (following the notations of Ref.~\cite{Sch04,GKA13}).
It is precisely the fact that this probability involves a permanent, notoriously hard to compute \cite{Val79}, that makes the problem seemingly intractable on a classical computer. 

So far, the hardness of \textsc{BosonSampling} for classical computers is only formally established in the exact version of the problem, i.e. when one is asked to sample exactly from the output distribution described in Eq.~\ref{proba_permanent}. 
%The status of the approximate version is less clear, and is connected to a conjecture about the permanent of random Gaussian matrices: if one can prove a strong enough anti-concentration bound for this permanent, then \textsc{BosonSampling} will remain hard classically, provided the sampled distribution is close enough to the \textsc{BosonSampling} one in trace distance \cite{AA13}.
The status of the approximate version is less clear and depends on two conjectures which, if established, would imply that \textsc{BosonSampling} will remain hard classically, provided that the sampled distribution is close enough to the \textsc{BosonSampling} one in trace distance.
The first conjecture is that the permanent of random Gaussian matrices is not too concentrated around its expected value. The second one is the ``Permanent of Gaussians Conjecture'' which says that  approximating the permanent of a Gaussian matrix is a $\sharp\mathrm{P}$-complete problem (see Ref.~\cite{AA13} for details). 
 In fact, this approximate version of the problem is of paramount importance since an experimental implementation cannot be expected to sample from the exact distribution. 
Indeed, many sources of noise or errors might degrade the quality of the experimental sampling. 
In Ref.~\cite{AA13}, five such sources of errors are mentioned: $(i)$ the single photons might not be perfectly indistinguishable; $(ii)$ the beamsplitters and phase shifters might not implement exactly the desired transformation; $(iii)$ the optical network might be subject to losses; $(iv)$ the detectors might not have a perfect efficiency and $(v)$ various synchronization issues might degrade the indistinguishability of the single photons even further.
In particular, it is worth noting that the effect of losses, imperfect detection and imperfect sources sending vacuum with a nonzero probability can be treated similarly (provided that these imperfections are invariant under a permutation of the modes). Indeed, all such imperfections can be modeled as effective losses, acting independently on each mode and commuting with the optical network. According to Refs \cite{RR12,Roh12}, the lossy variant of \textsc{BosonSampling} remains a hard problem classically, provided the losses are not too large: one can then use postselection to recover the original problem, and this postselection can be implemented efficiency is the number of single photons is on the order of 20 to 30. 

The only imperfection that has not been addressed so far deals with the quality of the implementation of the desired unitary transformation: given an instance of the \textsc{BosonSampling} problem characterized by a unitary matrix $U^\mathrm{BS}$, how accurate should the implementation of $U^\mathrm{BS}$ be in order for the sampled distribution to be reasonably close to the ideal one? 
If the calibration of the elementary gates (here beamsplitters and phase-shifters) is too poor, each elementary gate will actually implement a transformation slightly different from the ideal one. These errors will accumulate, and as the numbers of modes and photons increase, such small deviations will lead to an output distribution far from the ideal one.
This problem is particularly relevant for several reasons. First, contrary to \textsc{Factoring}, \textsc{BosonSampling} is not believed to be in the complexity class NP, meaning that there is no efficient way to verify that the experimental samples really correspond to a \textsc{BosonSampling} of $U^\mathrm{BS}$ (see however Ref.~\cite{ABM12}). Therefore, it would be useful to know in advance whether a given experimental implementation will lead or not to meaningful results, for a given choice of parameters $n$ and $m$. 
A second point is that one of the main advantages of \textsc{BosonSampling} over other problems is that it does not seem to require a fault-tolerant quantum computer for its implementation. At least, this is the hope in the regime of $n$ and $m$ where classical simulation of \textsc{BosonSampling} becomes difficult.

In this paper, we study this effect in details. 
In particular, the faulty implementation of the elementary unitaries will be our only source of imperfection: we therefore assume that the experimentalist has access to perfect single photons sources, perfect detectors, and that the optical network is lossless. Although unrealistic, this assumption is useful in order to clearly analyze the effect of the imprecise implementation of the beamsplitters or phase-shifters.

Our main result will be that in order for the implementation to provide a reasonably good output distribution, the average fidelity of the elementary gates should scale at least like $1 - O(1/n^2)$. 
This result indicates that the faulty implementation might not be the main experimental concern for implementations where $n \approx 30$, which should be sufficient to establish quantum supremacy. One should not forget, however, that only one single source of noise is taken into account here.

The outline of the paper is as follows. We first state the problem more precisely in Section \ref{statement} and explain our model of noise in Section \ref{model-noise}. Then, in Section \ref{walk}, we study a matrix-valued random walk corresponding to the propagation of noise in the optical network. In Section \ref{envelope}, we give a back of the envelope estimate for the effect of the noise on the sampling, and the following two sections aim at proving rigorously this result.

\section{Statement of the problem}
\label{statement}

Our goal is to study how well a basic experimental implementation of Boson Sampling can perform. Let us therefore consider an instance of \textsc{BosonSampling} characterized by a unitary $U^\mathrm{BS}$ acting on $\mathbbm{C}^m$ drawn from the Haar measure and a number $n$ of photons.
Let us denote by $p$ the ideal output probability distribution corresponding to this instance, that is the one we want to sample from, and by $\tilde{p}$ the experimental probability distribution. In this work, as we mentioned, we wish to focus on one specific source of imperfection, namely the unavoidable imprecision in the implementation of the elementary gates (beamsplitters and phase-shifters), assuming therefore that the rest of the experiment is ideal. This means that we consider that the input state is precisely $|1\rangle^{\otimes n} |0\rangle^{\otimes (m-n)}$, that the photon counters have perfect efficiency and zero dark count rate, and that the optical circuit is lossless. 

In order to assess the quality of the sample provided by the experiment, we compute the trace distance between the two probability distributions, $\|p - \tilde{p}\|_1$, which has the nice operational interpretation that the maximal probability of correctly distinguishing $p$ from $\tilde{p}$ when given one sample of either one, is precisely $\frac{1}{2}(1 + \frac{1}{2} \|p - \tilde{p}\|_1)$.
With this measure, we will say that the implementation provides the correct output if $\|p-\tilde{p}\| \ll 1$, that is, if the experimental distribution is almost indistinguishable from the ideal one. 

Ideally, we would like to get a bound on $\|p-\tilde{p}\|$ for any choice of unitary $U^\mathrm{BS}$. Unfortunately, this problem seems quite challenging, and we will therefore use the following trick. Instead of comparing the ideal implementation of $U^\mathrm{BS}$ with its realistic implementation which we will denote by $\Phi(U^\mathrm{BS})$ (where $\Phi$ should be understood as a random endomorphism of the unitary group $U(m)$), we will compare the ideal implementation of $U^\mathrm{BS}$, \emph{followed by} its inverse $U^{\mathrm{BS} \dagger}$, that is the identity map over $\mathbbm{C}^m$ , to its implementation $\Phi(U^\mathrm{BS}) \cdot \Phi(U^{\mathrm{BS} \dagger})$. 
For a reasonable model of noise, the map $\Phi$ acts identically and independently on each optical element, meaning that the implementation $\Phi(U^{\mathrm{BS} \dagger})$ of the inverse of $U^\mathrm{BS}$ will not correspond to the inverse of the implementation of $U^\mathrm{BS}$, that is $\Phi(U^{\mathrm{BS} \dagger}) \Phi(U^{\mathrm{BS} }) \ne \mathbbm{1}_m$.
Our idea is therefore to  compare the implementation of a random circuit followed by its inverse to the identity channel. Intuitively, this comparison provides a good prediction of how a Boson Sampling experiment will perform on a random instance of the problem.

 The main interest of this approach is that the distribution $p$ becomes particularly simple: the output sequence $\vec{s}_1=(1, \ldots, 1, 0, \ldots, 0)$ is assigned a probability 1, and all other events (corresponding to all the other ways to distribute $n$ photons in $m$ modes) have probability 0.
The distance between $p$ and $\tilde{p}$ is then very easy to express. Denoting by $\tilde{p}_1$ the probability of the output sequence $\vec{s}_1 = (1, \ldots, 1, 0, \ldots, 0)$ for the distribution $\tilde{p}$, the distance is given by $\|p - \tilde{p}\| = 2 (1- \tilde{p}_1)$. The probability $\tilde{p}_1$ can be computed thanks to Eq.~\ref{proba_permanent}: 
\begin{align}
\tilde{p}_1 = |\mathrm{Per}([\Phi(U^\mathrm{BS}) \cdot \Phi(U^{\mathrm{BS} \dagger})]_{n\times n})|^2,
\end{align}
 where $[A]_{n\times n}$ is the $n\times n$ matrix corresponding to the first $n$ rows and columns of $A$.
For a perfect implementation, this unitary is simply the identity $\mathbbm{1}_m$ and $\mathrm{Per}([\mathbbm{1}_m]_{n\times n}) = 1$. Therefore, in order to assess the quality of the implementation, one needs to see how $\tilde{p}_1$ deviates from 1.
In general, we expect the unitary transformation $\Phi(U^\mathrm{BS}) \cdot \Phi(U^{\mathrm{BS} \dagger})$ to remain close to the identity, and our goal will be to see how close, and more specifically, how close the permanent of its $n \times n$ upper-left minor is to 1. 

Finally, we note that there are two distinct sources of randomness in a Boson Sampling experiment. The first source is the choice of the unitary $U^\mathrm{BS}$ to be implemented. As we mentioned, the unitary is typically chosen according to the Haar measure on the unitary group $U(m)$. The second source of randomness is the noise $\Phi$ due to the imperfect implementation of $U^\mathrm{BS}$. Our goal will be to compute the expected distance $\mathbbm{E}_\mathrm{BS} \mathbbm{E}_\Phi \,\|p - \tilde{p}\|$ when averaging over these two sources of noise, and see how this measure of quality behaves as a function of the level of noise and the size of the experiment. Here, $\mathbbm{E}_\mathrm{BS}$ refers to the expectation over the choice of unitary $U^\mathrm{BS}$ and $\mathbbm{E}_\Phi$ refers to the expectation over the noise in the implementation. 

In particular, a proposal of experimental Boson Sampling only makes sense if it is possible to bring the noise level sufficiently down to ensure that the quantity $\mathbbm{E}_\mathrm{BS} \mathbbm{E}_\Phi \,\|p - \tilde{p}\|$ does not become too large with increasing numbers of modes and photons. 

\section{Our model of noise}
\label{model-noise}

Choosing the right model for the noise is crucial, but nontrivial. A good model should display at least two features: it should allow us to compute relevant quantities, and even if it is only an approximation, it should predict the essential properties of an imperfect implementation of Boson Sampling. More specifically, it should allow us to see how the distance $\mathbbm{E}_\mathrm{BS} \mathbbm{E}_\Phi \|p - \tilde{p}\|$ will depend on the parameters $n$ (number of single photons) and $m$ (number of modes) of the problem. 

\begin{figure}
\centering
 \includegraphics[width=90mm]{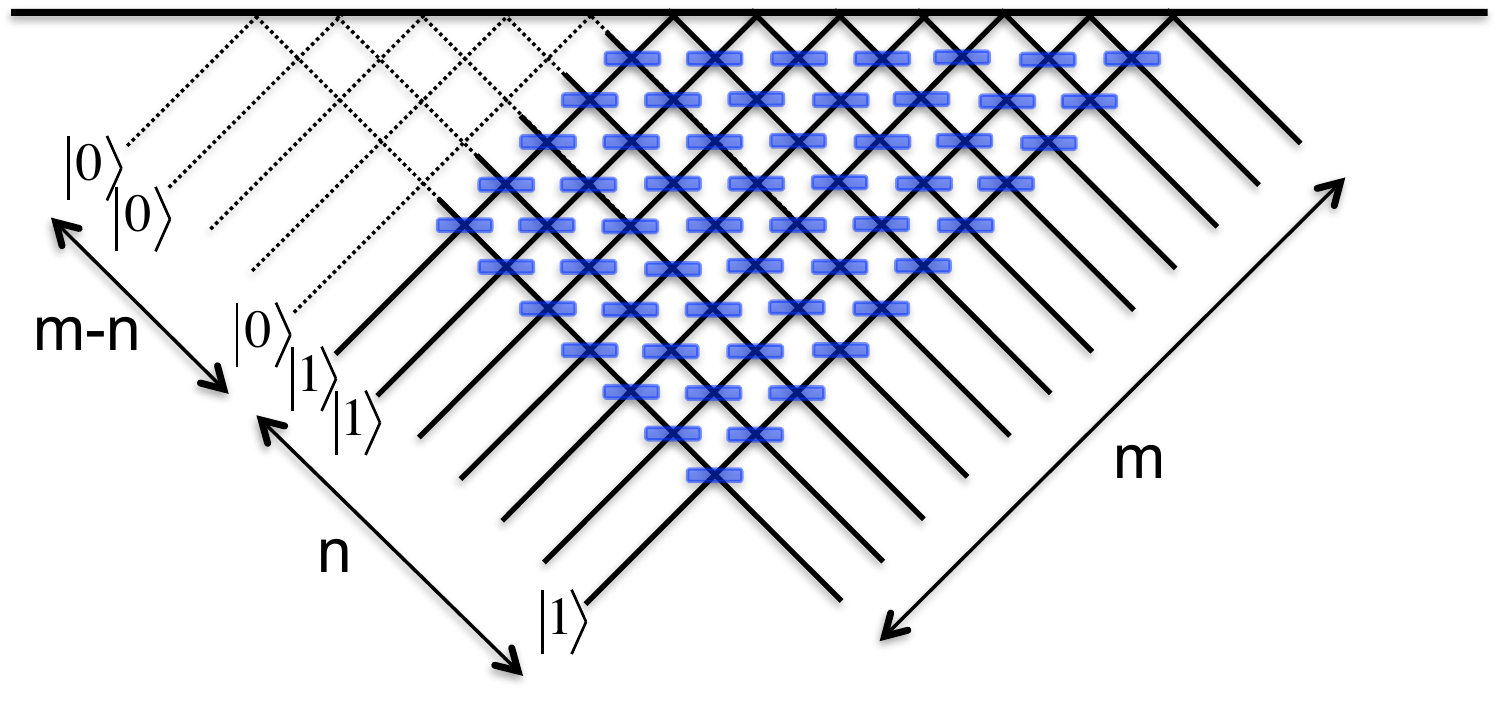}
  \fcaption{\textbf{Implementation of the Boson Sampling experiment, following the scheme of Ref.~\cite{RZB94}.} Input modes are on the left. The line at the top of the figure depicts a mirror.  The upper left beamsplitters only act on vacuum modes and can be ignored. 
 The remaining beamsplitters (in blue) are considered to be noisy in this work.
  }
   \label{figure:network}
\end{figure}

First, let us describe how a given unitary is implemented. A general algorithm due to Reck et al \cite{RZB94} provides a way to implement any unitary $U^{\mathrm{BS}}$ acting on $m$ modes via a network of beamsplitters and phase-shifters. Without loss of generality, the phase-shifters can be absorbed within the beamsplitters (which are now described by complex unitaries in $U(2)$), and the network is arranged in the way depicted on Fig.~\ref{fig:scheme}.
For a general unitary, the number of beamsplitters required is $\tbinom{m}{2}$ if the number of modes is $m$. 
However, in a Boson Sampling experiment, most of the input modes contain vacuum, and the beamsplitters only acting on vacuum will not play any role in the computation. These beamsplitters can be removed, thereby considerably simplifying the experimental scheme which only contains $N = n(m-(n+1)/2)$ beamsplitters (see Fig.~\ref{figure:network}).
We note now that our analysis relies on the specific implementation of the unitary described by Reck et al. However, we believe that alternative implementations would lead to similar results. 
Denoting by $U_k^{\mathrm{BS}}$ the $k^\mathrm{th}$ elementary gate to be applied, one ends up with a decomposition of $U^{\mathrm{BS}}$ given by $U^{\mathrm{BS}} = U_N^{\mathrm{BS}} U_{N-1}^{\mathrm{BS}} \cdots U_2^{\mathrm{BS}} U_1^{\mathrm{BS}}$.

Let us now describe our model of error for the implementation of Boson Sampling. We write $\Phi(U_k^{\mathrm{BS}})$ for the unitary transformation which is actually implemented, instead of $U_k^\mathrm{BS}$. Here $\Phi$ is a random map from $U(m)$ to $U(m)$ that acts independently on each elementary unitary. Moreover, we assume that if $U_k^\mathrm{BS}$ acts nontrivially on two modes, $k_1$ and $k_2$, then the unitary $\Phi(U_k^{\mathrm{BS}})$ will only act nontrivially on the same two modes, $k_1$ and $k_2$. Note that the independence assumption implies that in general, for a given unitary $U$, the unitaries $\Phi(U)$ and $\Phi(U^\dagger)$ are not inverse from each other, although this is approximately the case if the noise is sufficiently small. In fact, let us consider the matrix $\Phi(U_k^{\mathrm{BS} \dagger})\Phi(U_k^{\mathrm{BS}})$. Since this matrix is unitary, it can be written as $\exp(i \epsilon h_k)=\Phi(U_k^{\mathrm{BS} \dagger})\Phi(U_k^{\mathrm{BS}})$, where $\epsilon > 0$ is a parameter controlling the noise intensity, and $h_k$ is a Hermitian matrix. 
Our model of noise is to take $h_k$ acting nontrivially only on the same two modes as $U_k^\mathrm{BS}$, and the restriction of $h_k$ to these two modes to be a $2 \times 2$ random matrix drawn from the Gaussian Unitary Ensemble. This means that the four non-zero entries $\left(\begin{smallmatrix} \alpha_k & \gamma_k \\ \gamma_k^* & \beta_k \end{smallmatrix}\right)$ are such that $\alpha_k, \beta_k \sim \mathcal{N}(0,1)_{\mathbbm{R}}$ and $\gamma_k \sim \mathcal{N}(0,1/2)_\mathbbm{C}$, where $\mathcal{N}(0,1)_\mathbbm{K}$ refers to the normal distribution with mean 0 and variance 1 over the field $\mathbbm{K}$. 

Let us see how the intensity of the noise is linked to the average fidelity $F$ of each beamsplitter. This average fidelity is computed as the fidelity of the one-qubit gate characterized by the same $2\times 2$ unitary as the beamsplitter, and it corresponds to the fidelity of a one-qubit gate for a dual-rail encoding:
\begin{align}
F &= \mathbbm{E}_\Phi \int_{\phi \in \mathbbm{C}^2} \langle \phi |\Phi(U_k^{\mathrm{BS} \dagger})\Phi(U_k^{\mathrm{BS}}) |\phi\rangle d\phi\\
 &= \mathbbm{E}_\Phi  \int_\phi \langle \phi |\exp(i \epsilon h_k) |\phi\rangle d\phi \\
&=\mathbbm{E}_\Phi  \mathrm{tr}\left[\int_\phi |\phi\rangle \langle \phi| d\phi  \exp(i \epsilon h_k)  \right]\\
&\approx \frac{1}{2} \mathrm{tr} \left[\mathbbm{1}_2 \left(\mathbbm{1}_2- \epsilon^2 \left( \begin{smallmatrix} 1&1\\1&1 \end{smallmatrix} \right)\right)\right]\\
& = 1-\epsilon^2.
\end{align}

Let us write $V_k = \Phi(U_k^\mathrm{BS})$ so that $\Phi(U_k^{\mathrm{BS} \dagger}) =  e^{i \epsilon h_k} V_k^\dagger$. We note that the matrix $V_k$ follows approximately the same distribution as the matrix $U_k^\mathrm{BS}$ as long as the noise parameter $\epsilon$ remains sufficiently small. 
The implementation of the identity $\mathbbm{1}_m = U^\mathrm{BS}\cdot U^{\mathrm{BS} \dagger}$ then reads:
\begin{align*}
\Phi(U^{\mathrm{BS}} \cdot U^{\mathrm{BS} \dagger} ) &= \Phi(U_N^{\mathrm{BS}}) \Phi(U_{N-1}^{\mathrm{BS}}) \cdots \Phi(U_2^{\mathrm{BS}}) \Phi(U_1^{\mathrm{BS}}) \Phi( U_1^{\mathrm{BS}^\dagger})  \Phi( U_2^{\mathrm{BS}^\dagger}) \cdots \Phi(U_N^{\mathrm{BS} \dagger} ) \\
&= \Phi(U_N^{\mathrm{BS}}) \Phi(U_{N-1}^{\mathrm{BS}}) \cdots \Phi(U_2^{\mathrm{BS}})  V_1 e^{i \epsilon h_1} V_1^\dagger \Phi( U_2^{\mathrm{BS}^\dagger}) \cdots \Phi(U_N^{\mathrm{BS} \dagger} ) \\
&= \Phi(U_N^{\mathrm{BS}}) \Phi(U_{N-1}^{\mathrm{BS}}) \cdots V_2  V_1 e^{i \epsilon h_1} V_1^\dagger  e^{i \epsilon h_2}  V_2^\dagger  \cdots \Phi(U_N^{\mathrm{BS} \dagger} )
\end{align*}
Let us now introduce the following sequence $(H_k)$ of random $m \times m$ hermitian matrices defined by:
\begin{align}
\label{recurrence}
H_1 = h_1; \quad \exp(i \epsilon H_{k+1}) =V_{k}  \exp(i \epsilon H_{k}) V_{k}^\dagger e^{i \epsilon h_{k+1}} .
\end{align}
With this notation, the $N^\mathrm{th}$ element of the sequence is exactly what we wish to study since $\exp(i \epsilon H_N) =  \Phi(U^{\mathrm{BS}} \cdot U^{\mathrm{BS} \dagger} )$.
In particular, we are interested in the expected value of the permanent of the $n\times n$ upper left minor of the matrix $\exp(i \epsilon H_N)$.

Before attacking this problem, we summarize our model for Boson Sampling. First, a unitary matrix $U^{\mathrm{BS}}$ acting on $\mathbbm{C}^m$ is drawn from the Haar measure. This unitary is decomposed into an array of (complex) beamsplitters arranged as depicted on Fig.~\ref{fig:scheme} following the scheme of Reck et al \cite{RZB94}.
Then, we assume that for each optical element, a beamsplitter $\Phi(U_k^\mathrm{BS})$ is implemented instead of the ideal one, $U_k^{\mathrm{BS}}$. The random map $\Phi$ acts identically and independently on each beamsplitter, and we model the noise in such a way that 
$$  \Phi(U_k^\mathrm{BS})\Phi(U_k^{\mathrm{BS} \dagger}) = \exp(i \epsilon h_k),$$
where $\epsilon >0$ is a parameter controlling the intensity of the noise (and related to the average fidelity of each beamsplitter through $F \approx 1 - \epsilon^2$) and $h_k$ is a random hermitian matrix drawn from the Gaussian Unitary Ensemble, and acting non trivially only on the same two modes as $U_k^\mathrm{BS}$.
In the following, we will limit our analysis to the regime where $n \epsilon \ll 1$ and prove that $\mathbbm{E}_\mathrm{BS} \mathbbm{E}_\Phi \,\|p - \tilde{p}\| = \Omega( n^2 \epsilon^2)$ in this regime.

\section{Study of the matrix-valued random walk $(H_k)$}
\label{walk}
 
 In order to compute $\mathbbm{E}_\mathrm{BS} \mathbbm{E}_\Phi \,\|p - \tilde{p}\|$, we proceed in two main steps. First, we need to study the behavior of the random matrix $H_N$, and then we will compute the permanent of its $n \times n$ upper-left minor. 
 In this section, we study the random walk $(H_k)_{k=1 \cdots N}$ over Hermitian matrices. 
 
 The recurrence relation of Eq.~\ref{recurrence} can be rewritten as
 \begin{align}
  \exp (i\epsilon H_{k+1}) = \exp (i\epsilon V_{k} H_{k} V_{k}^\dagger) \exp (i \epsilon h_{k+1}).
  \end{align}
Since all the matrices in the exponent are small, one is tempted to use the Baker-Campbell-Hausdorff formula, which states that for arbitrary matrices $X$ and $Y$,
\begin{align}
\label{BCH}
\ln (\exp X \exp Y) = X + Y + \frac{1}{2} [X,Y] + \frac{1}{12} [X, [X,Y]] - \frac{1}{12} [Y, [X,Y]] + \cdots
\end{align}
and to discard all but the first two terms of the sum. 
 Certainly, this approximation holds for small values of the index $k$ since both $\epsilon V_{k} H_{k} V_{k}^\dagger$ and $\epsilon h_{k+1}$ are small matrices (for any choice of norm).
 As long as the approximation holds, one obtains
 \begin{align}
 \label{BCH_approx}
 H_{k} = V_k^\dagger \left( \sum_{i=1}^k V_k \cdots V_i h_i V_i^\dagger \cdots V_k^\dagger\right) V_k.
 \end{align}
 The question is then to determine for how long this approximation holds. 
 We will prove in Appendix \ref{approx_BCH} that in the regime where $n \epsilon \ll 1$, one can indeed safely neglect second and higher order terms in the expansion of the Baker-Campbell-Haussdorff formula.  
 Therefore we consider that Eq.~\ref{BCH_approx} holds for all values of $k$ up to $N$. Moreover, in order to slightly simplify the notation, we forget about the last conjugation with the unitary $V_N$ and take $H_N$ to be equal to:
 \begin{align}
 H_{N} =  \sum_{i=1}^N V_N \cdots V_i h_i V_i^\dagger \cdots V_N^\dagger
 \end{align}
 where $N = O(nm)$ is the total number of beamsplitters in the implementation of Boson Sampling. Note that forgetting one conjugation does not affect our results too much in the regime where $N$ is large.

\section{Computing the permanent, a back-of-the-envelope estimate}
\label{envelope}

The permanent of a matrix is notoriously much more difficult to handle than its cousin, the determinant. While this is of course one of the reasons why we expect \textsc{BosonSampling} to be intractable on classical computer, it certainly makes our task harder when trying to analyze how the permanent of the $n\times n$ minor of $\exp(i \epsilon H_N)$ will behave when varying the parameters $n$ and $m$ of the Boson Sampling experiment. In order to avoid cluttering the notation, we will denote by $V = (v_{i,j})$ the $n\times n$ upper-left minor of $\exp(i \epsilon H_N)$ so that the the quantity we wish to compute is $\mathbbm{E}_{\mathrm{BS}} \mathbbm{E}_\Phi \mathrm{Per}(V)$.

Following the notation of Appendix B of Ref.~\cite{AA13}, given a vector $x \in \mathbbm{C}^n$, we define the polynomial
\begin{align}
\label{def_rys}
\mathrm{Rys}_x (V) := n^n \prod_{i=1}^n x_i \sum_{j=1}^n v_{i,j} x_j,
\end{align}
where the name $\mathrm{Rys}$ refers to a result of Ryser stating that $\mathrm{Per}\, V = \mathbbm{E}_{x} \mathrm{Rys}_x (V)$, where the expectation is over the uniform distribution for the vector $x \in \{-1/\sqrt{n},1/\sqrt{n}\}^n$.

Before delving into the computation, it is useful to better understand the structure of the matrix $V$, and to see what to expect for a typical value of $\mathrm{Per}\, V$. We insist that the following argument is in no way rigorous, but it will give us the correct behavior of  $\mathrm{Per}\, V$, as we will confirm later. 
Let us first expand the exponential:
\begin{align}
\exp(i\epsilon H_N) = \mathbbm{1}_m + i \epsilon H_N -\frac{\epsilon^2}{2} H_N^2 + \cdots
\end{align}
We now wish to understand the behavior of the random variable corresponding to $(H_N)_{i,j}$. From a symmetry argument, we expect it to be centered. 
We also expect that all the entries of the matrix $H_N$ should be approximately identically distributed. 
From that, we obtain that $\mathrm{Var} (H_N)_{i,j} \approx \frac{1}{m^2} \mathbbm{E} \|H_N\|^2_\mathrm{HS}$, where $\|\cdot\|_\mathrm{HS}$ refers to the Hilbert-Schmidt norm. 
Using the independence of the matrices $h_i$,
\begin{align}
\mathbbm{E}_\Phi  \|H_N\|^2_\mathrm{HS} = \mathbbm{E}_\Phi \sum_{i=1}^N \|h_i\|_{\mathrm{HS}}^2 = O(N),
\end{align}
which gives $\mathrm{Var} (H_N)_{i,j} = O(N/m^2)$.

For small values of $\epsilon$, one expects that $v_{i,i} \approx 1 - \frac{\epsilon^2}{2} \sum_{j=1}^m |(H_N)_{i,j}|^2$ and $v_{i,j} \approx i \epsilon (H_N)_{i,j}$ for $i \ne j$. Moreover, since we expect $(H_N)_{i,j}$ to be a centered random variable of variance $O(N/m^2)$, we infer that $ \sum_{j=1}^m |(H_N)_{i,j}|^2$ should be on the order of $N/m$. In particular, $x_i \sum_{j=1}^n v_{i,j} x_j$ should remain close to $v_{i,i}/n$ and  $\mathrm{Rys}_x (V)$ be approximately equal to the product $\prod_{i=1}^n v_{i,i}$.
One therefore expects the permanent to be of the order of $\left(1-\frac{\epsilon^2 N}{2m}\right)^n \approx 1- \frac{Nn\epsilon^2}{2m}$. Recalling that $N = O(n m)$ and $m = O(n^2)$ leads to the prediction that the permanent should be of the order of $1 - O(n^2\epsilon^2)$.

In other words, this very hand-wavy argument tells us that we should expect 
\begin{align}
\mathbbm{E}_\mathrm{BS} \mathrm{E}_\Phi \|p - \tilde{p}\| = O(n^2 \epsilon^2) = O(n^2(1-F)),
\end{align}
where $F$ is the average fidelity by beamsplitter.
The rest of the paper aims at making this statement more rigorous. In particular, we will prove a lower bound on $\mathbbm{E}_\mathrm{BS} \mathrm{E}_\Phi \|p - \tilde{p}\|$.

\section{A bound on the permanent}

In this section, we relate the value of the permanent of $V$, which is too hard to analyze directly, to a quantity easier to compute corresponding roughly to the average singular value of $V$. 
Take a vector $x \in \mathbbm{C}^n$ such that $|x_i|=1/\sqrt{n}$ and define the vector $y = (y_1, \ldots, y_n)$ where $y_i = \sum_{j=1}^n v_{i,j} x_j$. Then the following sequence of (in)equalities holds:
\begin{align}
|\mathrm{Per}\, V | &= n^{n} \left|\mathbbm{E}_x  \prod_{i=1}^n x_i y_i \right|\\
& \leq n^{n} \mathbbm{E}_x \left| \prod_{i=1}^n x_i y_i \right|\\
&=n^{n/2}\mathbbm{E}_x  \left| y_1 \cdots y_n \right|\\
& \leq n^{n/2}\mathbbm{E}_x  \left( \frac{1}{n} \sum_{i=1}^n |y_i|\right)^n \label{arithmetic-geometric} \\
& = n^{n/2} \mathbbm{E}_x\left( \frac{1}{n} \left[ \sum_{i=1}^n |y_i|^2  - \frac{1}{n} \sum_{i < j} (|y_i|-|y_j|)^2 \right]\right) ^{n/2} \label{nontight} \\
& \leq \mathbbm{E}_x\|y\|^n   \label{final_norm}
\end{align}
where in the first equality, the expectation is over the uniform measure over complex numbers of norm $\frac{1}{\sqrt{n}}$ for each $x_i$. Line \ref{arithmetic-geometric} follows from the arithmetic-geometric inequality. Note that Eq.~\ref{final_norm} can be obtained directly from Eq.~\ref{arithmetic-geometric} by applying Cauchy-Schwarz to the vectors $(1/n, \ldots, 1/n)$ and $(|y_1|, \ldots, |y_n|)$.
Squaring the final inequality, and using the convexity of the function $(x \mapsto x^2)$ gives us an upper bound on the square of the permanent: 
\begin{align}
|\mathrm{Per}\, V |^2 \leq \mathbbm{E}_x\|y\|^{2n}.
\end{align}
In order to bound the value of the permanent away from 1, we will show that $\|y\|$ is bounded away from one with non negligible probability\footnote{We should ideally exploit the various points where the inequalities above are not tight. 
Unfortunately, the analysis of terms such as $\sum_{i < j}  (|y_i|-|y_j|)^2$ in Eq.~\ref{nontight} is not easy to perform. 
An obvious path to improve on our results and get a tighter bound would be to provide a better analysis of the various sources of loss in the inequalities above.}.

Let us slightly abuse notation, and also denote by $x$ the vector $(x_1, \ldots, x_n, 0, \ldots, 0) \in \mathbbm{C}^m$. 
We also introduce $\Pi_n = \mathrm{diag}(1,\ldots, 1, 0, \ldots, 0)$, the projector on the $n$-dimensional subspace of $\mathbbm{C}^m$ spanned by the first $n$ vectors of the canonical basis. 
We also write $y$ for the $m$-dimensional vector $V x = \Pi_n \exp(i \epsilon H_N) \Pi_n x$. 

Since the matrix $\exp(i \epsilon H_N)$ is unitary, 
\begin{align}
1 &= \| \exp(i\epsilon H_N) \Pi_nx \|^2 \\\
&= \| \Pi_n \exp(i\epsilon H_N)\Pi_n x \|^2 + \|(\mathbbm{1}_m - \Pi_n) \exp(i\epsilon H_N)\Pi_nx\|^2 \\
&= \|y\|^2 + \|(\mathbbm{1}_m - \Pi_n) \exp(i\epsilon H_N) \Pi_n x \|^2 \\
&\approx \|y\|^2 +  \|(\mathbbm{1}_m - \Pi_n) (\mathbbm{1}_m + i \epsilon H_N) \Pi_n x\|^2  \label{substitution}\\
&= \|y\|^2 + \epsilon^2 \|(\mathbbm{1}_m - \Pi_n) H_N \Pi_n x\|^2
\end{align}
where, in Eq.~\ref{substitution}, we neglected terms of order 2 or higher in the expansion of $\exp(i \epsilon H_N) \approx \mathbbm{1}_m + i\epsilon H_N$.
Indeed, to see that this is justified, we compute for instance the term of order 2, whose contribution is $\frac{\epsilon^4}{4} \|(\mathbbm{1}_m - \Pi_n) H_N^2 \Pi_n x\|^2$. We will prove in Appendix \ref{chernoff} that the spectral norm of $H_N$ is of order $O(\sqrt{n \ln n})$ which means that the term of order 2 is upper-bounded by $O(\epsilon^4 n^2 (\ln n)^2)$, which can indeed be neglected with respect to the first order term $O(\epsilon^2 n \ln n)$ in the regime of $\epsilon \ll 1/n$. 
Therefore, one obtains the bound:
\begin{align}
\mathbbm{E}_{\mathrm{BS}} \mathbbm{E}_\Phi |\mathrm{Per}\, V |^2 \leq \mathbbm{E}_{\mathrm{BS}} \mathbbm{E}_\Phi \mathbbm{E}_x \left(1-\epsilon^2\|(\mathbbm{1}_m - \Pi_n) H_N \Pi_n x\|^2 \right)^n.
\label{bound_HN}
\end{align}

Let us denote by $X$ the squared-norm $\|(\mathbbm{1}_m - \Pi_n) H_N \Pi_n x\|^2$ so that we wish to compute $ \mathbbm{E}_{\mathrm{BS}} \mathbbm{E}_\Phi \mathbbm{E}_x \left(1-\epsilon^2 X \right)^n$.
In order to obtain an upper bound on this quantity, we will compute two things: the expectation $\mathbbm{E} X$ (where the expectation is over all 3 sources of randomness) together with a bound of the type $\mathbbm{P}[X \geq X_\mathrm{max}(\delta)]\leq \delta$, where $\delta$ will be an arbitrary small parameter, and $X_\mathrm{max}(\delta)$ will be a function increasing slowly as $\delta$ tends to zero. 
Our goal will be to show that the random variable $X$ is on the order of $\mathbbm{E}X$ with a non negligible probability.
Indeed, introducing $\mu = \mathbbm{E}X$ and $p_0 = \mathbbm{P}\left[X \geq \mathbbm{E}X/2 \right]$ one can then use the trivial upper bound
\begin{align}
\mathbbm{E} (1-\epsilon^2 X)^n &\leq \left(1-\frac{\epsilon^2 \mu}{2}\right)^n \mathbbm{P}[X \geq \mu/2] + 1\times  \mathbbm{P}[X \leq \mu/2]\\
& =  p_0\left(1-\frac{\epsilon^2 \mu}{2}\right)^n + 1- p_0
\end{align}
where we used that $(1-\epsilon^2 X)^n \leq 1$. 
Recall that for $\delta \leq 1/n$, the following inequality can be proved with standard functional analysis:
\begin{align}
(1-\delta)^n \leq 1 - (1-e^{-1}) n \delta.
\end{align}
Since $\epsilon^2 \mu/2 \leq 1/n$ in the regime of $\epsilon$ that we consider, we infer that 
\begin{align}
\label{final-bound}
\mathbbm{E} (1-\epsilon^2 X)^n \leq 1 - \frac{1}{2}(1-e^{-1})n p_0 \epsilon^2 \mu.
\end{align}
Our initial objective was to bound the trace distance between the true output distribution $p$ of \textsc{BosonSampling}, and the experimental distribution $\tilde{p}$. Our bound gives:
\begin{align}
\mathbbm{E}_{\mathrm{BS}}\mathbbm{E}_\Phi \|p- \tilde{p}\| \geq \left(1-\frac{1}{e}\right) n p_0\mu \epsilon^2.
\end{align}
The next two sections are devoted to computing $\mu = \mathbbm{E}X$ and $p_0 = \mathbbm{P}\left[X \geq \mathbbm{E}X/2 \right]$.

\section{Analysis of $\mu = \mathbbm{E} \|(\mathbbm{1}_m - \Pi_n) H_N \Pi_n x\|^2$}

In order to analyze the norm of $(\mathbbm{1}_m - \Pi_n) H_N \Pi_n x$, we need, \emph{at last}, to look more carefully at the structure of the matrix $H_N$. Recall that $H_N = \sum_{k=1}^N g_k$, where the matrix $g_k$ is given by:
\begin{align}
g_k = V_N \cdots V_{k} h_k V_{k}^\dagger \cdots V_N^\dagger.
\end{align}
The independence of the matrices $h_k$ gives: 
\begin{align}
\mathbbm{E}_\Phi \|(\mathbbm{1}_m - \Pi_n) H_N \Pi_n x \|^2 &= \sum_{k=1}^N\mathbbm{E}_\Phi \|(\mathbbm{1}_m - \Pi_n) g_k \Pi_n x \|^2.
\end{align}
Let us consider the quantity
\begin{align}
\mathcal{N}_k &:=  \|(\mathbbm{1}_m - \Pi_n) g_k \Pi_n x \|^2 \\
&= \langle x | \Pi_n V_N \cdots V_{k}h_k  V_{k}^\dagger \cdots V_N^\dagger (1-\Pi_n) V_N \cdots V_{k} h_k V_{k}^\dagger \cdots V_N^\dagger \Pi_n |x\rangle
\end{align}
When taking the expectation over $\Phi$, we need to compute $\mathbbm{E}_\Phi h_k^{\otimes 2}$. It is easy to check that if $a, b \sim \mathcal{N}(0, 1)_\mathbbm{R}$ and $c \sim \mathcal{N}(0,1/2)_\mathbbm{C}$, then
\begin{align}
\mathbbm{E}_\Phi \left[\begin{matrix} a & c \\ c^* & b \\ \end{matrix} \right]^{\otimes 2} =
\mathbbm{E}_\Phi \left[\begin{matrix} a^2 & 0 & 0 & 0 \\ 0 & 0 & cc^* &0 \\ 0 & cc^* & 0 & 0 \\ 0 & 0 &0 & b^2 \end{matrix} \right] = \left[\begin{matrix} 1 & 0 & 0 & 0 \\ 0 & 0 & 1 &0 \\ 0 & 1 & 0 & 0 \\ 0 & 0 &0 & 1 \end{matrix} \right]
\end{align}
Denoting by $k_1$ and $k_2$ the two modes the matrix $V_k$ acts on, and by $|e_{k_1}\rangle$ and $|e_{k_2}\rangle$ the corresponding vectors in the canonical basis of $\mathbbm{C}^m$, one obtains:
\begin{align}
\mathbbm{E}_\Phi h_k^{\otimes 2} = |e_{k_1}\rangle \langle e_{k_1}| \otimes |e_{k_1}\rangle \langle e_{k_1}| + |e_{k_1}\rangle \langle e_{k_2}| \otimes |e_{k_2}\rangle \langle e_{k_1}| + |e_{k_2}\rangle \langle e_{k_1}| \otimes |e_{k_1}\rangle \langle e_{k_2}| + |e_{k_2}\rangle \langle e_{k_2}| \otimes |e_{k_2}\rangle \langle e_{k_2}| .
\end{align}
Let us also observe that $\mathbbm{E}_{x} \Pi_n |x\rangle \langle x| \Pi_n \mathrm{d} x = \frac{1}{n} \Pi_n$.
We introduce %the vector $|z\rangle := V_{k}^\dagger \cdots V_N^\dagger \Pi_n |x\rangle$ and 
the operator $P_n^k := V_{k}^\dagger \cdots V_N^\dagger \Pi_n V_N \cdots V_{k}$. Here the index $k$ is a remainder that $P_n^k$ depends on $k$.
This gives:
\begin{align}
\mathbbm{E}_\Phi \mathbbm{E}_{x} \mathcal{N}_k &= \frac{1}{n} \left( \langle e_{k_1} |P_n | e_{k_1} \rangle \langle e_{k_1} |\mathbbm{1}_m-P_n | e_{k_1} \rangle + \langle e_{k_1} |P_n | e_{k_1} \rangle \langle e_{k_2} |\mathbbm{1}_m-P_n | e_{k_2} \rangle \right.\\
& \left. \quad  + \langle e_{k_2} |P_n | e_{k_2} \rangle \langle e_{k_1} |\mathbbm{1}_m-P_n | e_{k_1} \rangle  + \langle e_{k_2} |P_n | e_{k_2} \rangle \langle e_{k_2} |\mathbbm{1}_m-P_n | e_{k_2} \rangle \right)\\
&= \frac{1}{n} \left( p_1^k + p_2^k\right)\left(2-p_1^k-p_2^k\right)
\end{align}
where we defined $p_i^k :=   \langle e_{k_i} |P_n^k | e_{k_i}\rangle$.

\begin{figure}
\centering
 \includegraphics[width=90mm]{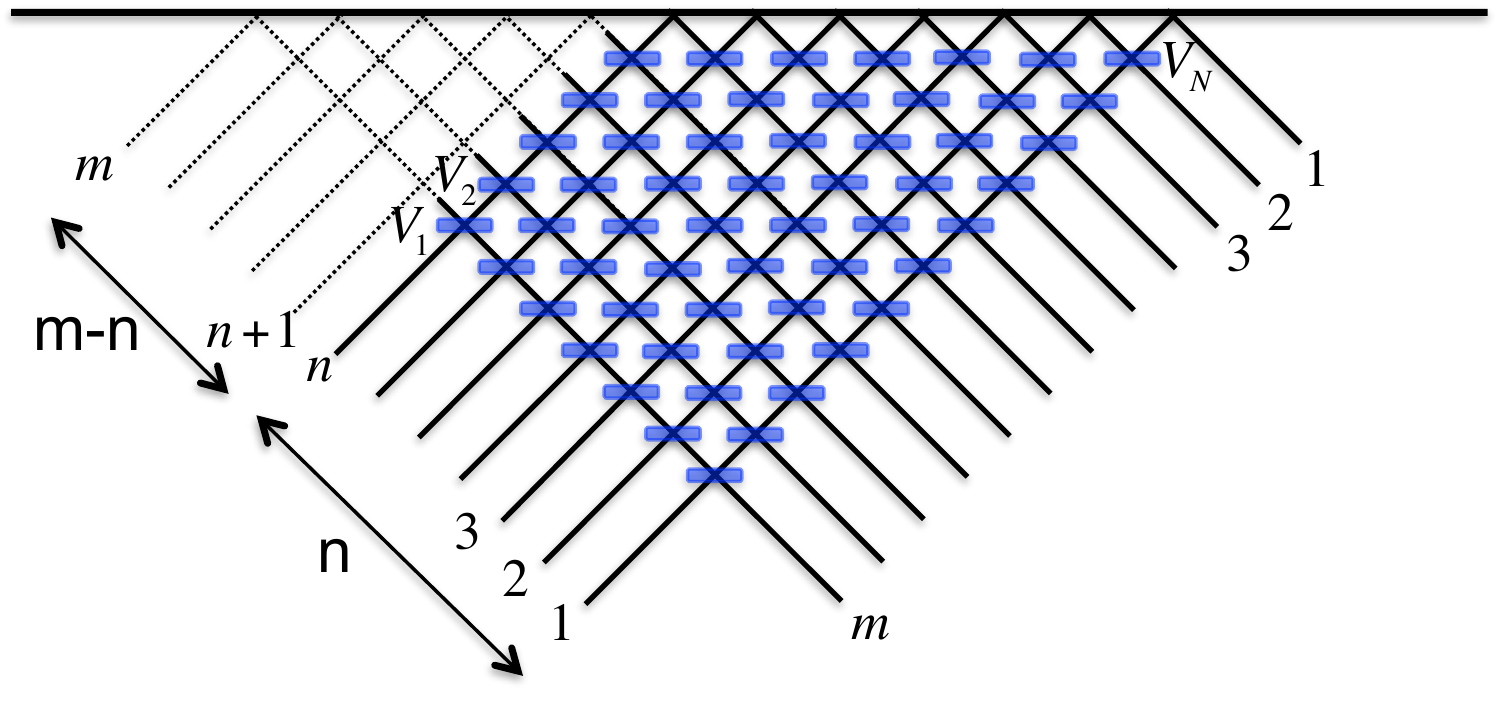}
  \fcaption{Each beamsplitter acts on two modes $i$ and $j$, which we denote by $\{i, j\}$: mode $i$ enters from the bottom left and mode $j$ from the top left. Input modes are labelled from $1$ to $m$ starting with 1 in the middle of the figure and ending with $m$ on the left of the figure.   The beamsplitters are labeled from bottom left to top right and from top left to bottom right. More precisely, $V_1$ acts on modes $n$ and $m$, so that $V_1 \equiv V_{\{n,m\}}$, then $V_2 \equiv V_{\{n, m-1\}}, \ldots, V_{m-n} \equiv V_{\{n, n+1\}}$. The final beamsplitter is $V_N \equiv V_{\{1,2\}}$.
  }
   \label{figure:network-numbers}
\end{figure}

In order to complete the calculations, we need to model more precisely how the unitaries $V_1, \ldots, V_N$ are chosen. For this, we consider the scheme of Reck \emph{et al.} from Ref.~\cite{RZB94} depicted on Fig.~\ref{figure:network-numbers}. However, we note that we do not expect our results to depend strongly on the specific choice of implementation of Boson Sampling. In this scheme, the various beamsplitters are not identically distributed, and in particular, it is not true that each beamsplitter is uniformly distributed on $U(2)$ (see Appendix \ref{reck} for details).

Consider the $k^\mathrm{th}$ beamsplitter acting on modes $k_1$ and $k_2$ with $1 \leq k_1 \leq n$ and $n+1 \leq k_2 \leq m$.
The quantity $p_1^k= \langle e_{k_1} |P_n | e_{k_1} \rangle $ is equal to the probability that given a single photon input in mode $k_1$ of a network of beamsplitters as in Fig.~\ref{figure:network-small}, the detector clicks in one of the first $n$ modes. 
We will show (in Appendix \ref{computation_p_1}) that $\mathbbm{E}_\mathrm{BS} \,p_1^k=\mathbbm{E}_\mathrm{BS} \, p_2^k = \frac{n}{k_2}$, and $\mathbbm{E}_\mathrm{BS} \, \left(p_1^k\right)^2=\mathbbm{E}_\mathrm{BS} \, \left(p_2^k\right)^2 = \frac{n(n+2)}{k_2(k_2+2)}$.
Moreover,  Cauchy-Schwarz gives:  $\mathbbm{E}_\mathrm{BS} \, p_1^k p_2^k \leq  \mathbbm{E}_\mathrm{BS} \, (p_1^k)^2$.
This leads to:
\begin{align}
\mathbbm{E}_\mathrm{BS} \mathbbm{E}_\Phi \mathbbm{E}_{x} \mathcal{N}_k \geq \frac{4}{n} \frac{n}{k_2} - \frac{4}{n}\frac{n(n+2)}{k_2(k_2+2)} =\frac{4(k_2-n)}{k_2(k_2+2)}.
\end{align}

\begin{figure}
\centering
 \includegraphics[width=80mm]{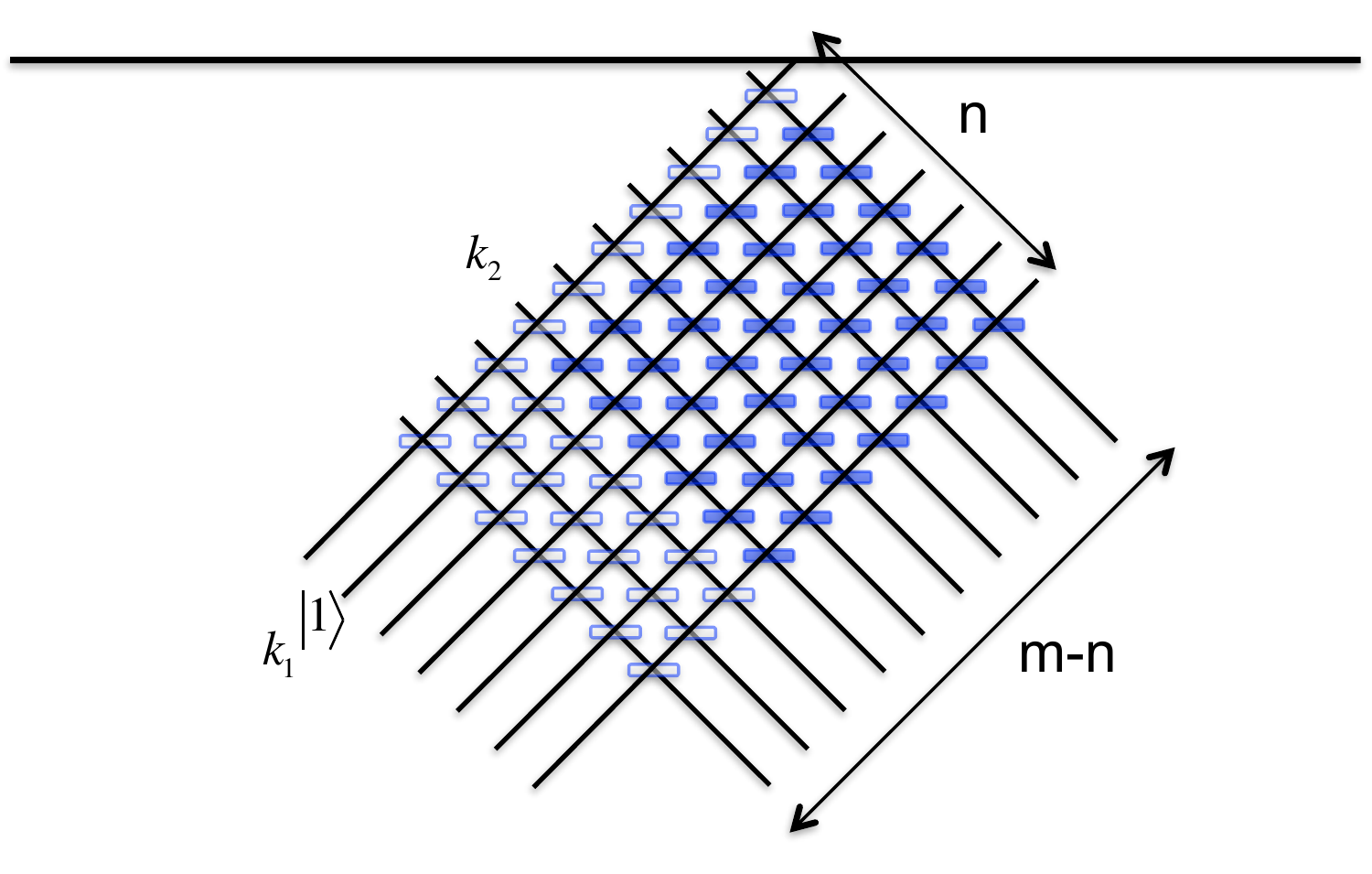}
  \fcaption{In order to compute $ \langle e_{k_1} |P_n | e_{k_1} \rangle $, one can consider a subset of the initial network of beamsplitters, and put a single photon in mode $k_1$. Only the beamsplitters acting on modes $\{i,j\}$ with $1 \leq i \leq k_1$ and $n \leq j \leq k_2$ need to be taken into account. The quantity $ \langle e_{k_1} |P_n | e_{k_1} \rangle $ is the probability that the photon is detected in one of the first $n$ output modes (on the right on the figure). }
   \label{figure:network-small}
\end{figure}

In the case where $1\leq k_1 \leq n$ and $k_1+1 \leq k_2 \leq n$, then the configuration of the network is incompatible with a photon exiting in a mode of index larger than $n$. This means that $p_i^k = 1$ and therefore that  $\mathbbm{E}_\mathrm{BS} \mathbbm{E}_\Phi \mathbbm{E}_{x} \mathcal{N}_k = 0$. 

Summing over all indices $k \in \{1, \ldots, N\}$ gives:
\begin{align}
\mu &=\mathbbm{E}_\mathrm{BS} \mathbbm{E}_\Phi \mathbbm{E}_{x} \|(\mathbbm{1}_m - \Pi_n) H_N \Pi_n x \|^2 \\
&= \sum_{k_2 =n+1}^m   \sum_{k_1=1}^n \mathbbm{E}_\mathrm{BS} \mathbbm{E}_\Phi \mathbbm{E}_{x} \mathcal{N}_k \\
&\geq 4n \sum_{k_2=n+1}^m \frac{k_2-n}{k_2(k_2+2)}\\
&= 4n \sum_{k_2=n+1}^m \frac{1}{k_2+2} - 4n^2 \sum_{k_2=n+1}^m \frac{1}{k_2(k_2+2)}\\
%& \geq 4n \sum_{k_2=n+1}^m \frac{1}{k_2+2} - 4n^2 \sum_{k_2=n+1}^m \frac{1}{k_2(k_2+1)}\\
& \geq 4n \sum_{k_2=n+1}^m \frac{1}{k_2+2} -  4n^2 \sum_{k_2=n+1}^m \left[\frac{1}{k_2} - \frac{1}{k_2+1} \right]\\
%& = 4n \sum_{k_2=n+1}^m \frac{1}{k_2+2} -  4n^2  \left[\frac{1}{n+1} - \frac{1}{m+1} \right]\\
& = 4n \sum_{k_2=1}^{m+2} \frac{1}{k_2} - 4n \sum_{k_2=1}^{n+2} \frac{1}{k_2} -  4n^2  \left[\frac{1}{n+1} - \frac{1}{m+1} \right]\\
&\geq 4n \left[ \ln(m+2) + \frac{1}{2m+9/2} - \ln (n+2) - \frac{1}{2n+4+1/3} \right]   -  4n^2  \left[\frac{1}{n+1} - \frac{1}{m+1} \right] \label{bound_harmonic}\\
%&\geq 4n \ln\left( \frac{m+2}{n+2}\right) - 4n \left[ \frac{1}{2n+4+1/3} + \frac{n}{n+1}\right] \\
%&\geq 4n \ln\left( \frac{m}{n}\right) -4n \ln\left(1+\frac{2}{n}\right) - 4n \left[ \frac{1}{2n+4+1/3} + \frac{n}{n+1}\right] \\
%&\geq 4n \ln\left( \frac{m}{n}\right) -8 - 4n \left[ \frac{1}{2n+4+1/3} + \frac{n}{n+1}\right] \\
&\geq 4n \ln\left( \frac{m}{e^2n}\right)
\end{align}
where inequality \ref{bound_harmonic} follows from the bounds of Ref.~\cite{GQ11} on the Harmonic number
\begin{align}
\frac{1}{2n+1/2} \leq \sum_{l=1}^n \frac{1}{l} - \ln n -\gamma \leq \frac{1}{2n+1/3}
\end{align}
where $\gamma = 0.577\ldots$ is Euler's constant.

\section{Bounding $p_0:= \mathbbm{P}\left[X \geq \mathbbm{E}X/2 \right]$}

Assume that the maximum possible value for $X$ is $M$ and that $\mathbbm{P}[X\geq X_\mathrm{max}(\delta)] \leq \delta$.
Let us express $\mu = \mathbbm{E}X$ as follows:
\begin{align}
\mu &= \int_{0}^{\mu/2} x \mathbbm{P}[X=x] dx + \int_{\mu/2}^{X_\mathrm{max}(\delta)} x \mathbbm{P}[X=x] dx + \int_{X_\mathrm{max}(\delta)}^{M} x \mathbbm{P}[X=x] dx \\
&\leq \frac{\mu}{2} (1-p_0) + p_0 X_\mathrm{max}(\delta) + \delta M
\end{align}
which gives
\begin{align}
p_0 \geq \frac{\mu- 2\delta M}{2X_\mathrm{max}(\delta)- \mu} \geq \frac{\mu- 2\delta M}{2 X_\mathrm{max}(\delta)}.
\end{align}

The quantity $M$ is given by $\pi^2 \times N \leq \pi^2 nm$ since each term of $H_N$ has a norm at most $\pi$. 
Let us choose $\delta =\frac{\mu}{4\pi^2 nm}$ so that $2\delta M \leq \mu/2$. For this value of $\delta$, we obtain $p_0 \geq \mu/(4 X_{\max}(\delta))$.

We prove in Appendix \ref{chernoff} that 
\begin{align}
\mathbbm{P}_{\Phi, \mathrm{BS}} \left[ \|(\mathbbm{1}_m -\Pi_n) H_N \Pi_N x \|^2 \geq  12 n \ln \frac{em}{n} \ln \frac{m}{\delta} \right] \leq \delta.
\end{align}
so that $X_\mathrm{max}(\delta) =  12 n \ln \frac{em}{n} \ln \frac{m}{\delta}$.

For our specific choice of $\delta$, this gives:
$$X_\mathrm{max}\left(\frac{\mu}{4\pi^2 nm}\right)\leq  12 n \ln \frac{em}{n} \ln \frac{4\pi^2nm^2}{\mu}  $$
and therefore 
\begin{align}
p_0 \geq \frac{\ln \frac{m}{e^2 n}}{12 \ln \frac{\pi^2 m^2}{\ln \frac{m}{e^2n}}}.
\end{align}

\section{Discussion and conclusion}

\begin{figure}
\centering
 \includegraphics[width=60mm]{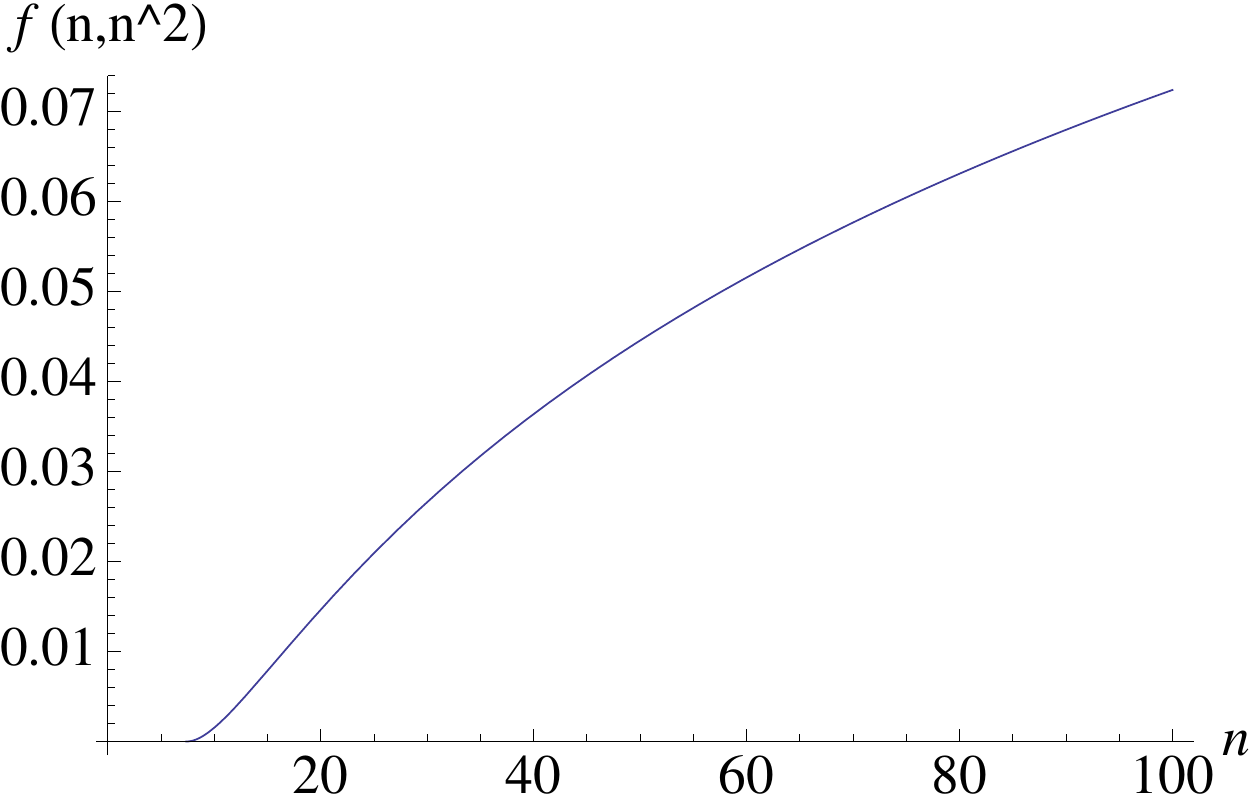}
  \fcaption{Function $f(n,m)$ evaluated for $m=n^2$. }
   \label{figure:f}
\end{figure}

Recalling the bound from Eq.~\ref{final-bound}, we obtain the following lower bound for the distance between the ideal probability distribution and the experimental one:
\begin{align}
\mathbbm{E}_{\mathrm{BS}}\mathbbm{E}_\Phi \|p- \tilde{p}\| &\geq f(n, m) n^2\epsilon^2 
\end{align}
where the function $f(n, m)$ is given by
\begin{align}
f(n, m) = \frac{\left(\ln \frac{m}{e^2 n} \right)^2}{10 \ln {\pi m} - 5\ln\ln \frac{m}{e^2 n}}.
\end{align}
One can check numerically that for $m = n^2$ and $n \geq 10$, then $f(n,m) \geq \left(\frac{\ln (n-10)}{20}\right)^2$.
The numerical value of $f$ is typically on the order of a few percent (see Fig.~\ref{figure:f}). 

This achieves our analysis of noise propagation in a Boson Sampling experiment: we have established a lower bound on the trace distance between the ideal and the experimental distributions, and we conjecture that the scaling is the right one in the regime where $n\epsilon \ll1$. Only the value of the constant might be modified by a more precise analysis. 
For reasonable values of $n$ and $m$, for instance $n \sim 30$ and $m \sim 1000$, one expects the impact of imperfect calibration of the beamsplitters and phase-shifters to remain negligible (i.e. on the order of a few percents) if the fidelity of each elementary gate is typically on the order of $0.999$. 
While certainly challenging, we do not expect such numbers to be a real problem in any forthcoming Boson Sampling experiment.

We end with a few open questions. First, we argued that our results are tight in the regime where $n\epsilon \ll 1$, but the analysis becomes much more involved when $n\epsilon$ is on the order of 1. Then, the approximation we made in the Baker-Campbell-Haussdorf formula is not valid anymore and our results become too optimistic. It would be very useful to derive an upper bound on $\mathbbm{E}_\mathrm{BS} \mathrm{E}_\Phi \|p - \tilde{p}\|$ to confirm that taking an average fidelity per gate of $1-O(1/n^2)$ is indeed sufficient to solve \textsc{BosonSampling}. 
Second, it would be highly relevant to be able to combine our analysis together with the effects of losses in the experimental setup. In fact, while some preliminary results have been obtained in Refs \cite{RR12, Roh12}, a rigorous study of the impact of losses on the hardness of \textsc{BosonSampling} is still missing.

\nonumsection{Acknowledgements}
\noindent
A.~L.~acknowledges financial support from the CNRS through the PEPS ICQ2013 CCQD, and R.~G.-P.~acknowledges financial support from the F.R.S.-FNRS.

\nonumsection{References}

\appendix 

\noindent

\section{Approximation in the Baker-Campbell-Haussdorf formula}
\label{approx_BCH}

In this appendix, we take the terms $[X,Y]$ into account in the Baker-Campbell-Haussdorf formula and prove that they do not modify significantly our results, provided that $n \epsilon \ll1$. More precisely, we include all second order terms in the computation of the random walk, and neglect third and higher order therms.
In fact, in the regime where $n \epsilon \approx 1$, these higher order terms cannot be neglected anymore.  
It is straightforward to check that if we only neglect third (and higher) order terms in the formula, after $N$ steps we end up with the following matrix $K_N$ instead of the matrix $H_N$ which is considered in the main text:
\begin{align}
K_N = H_N + \frac{\epsilon}{2} \sum_{1 \leq i < j \leq N} V_N \cdots V_j \left[ V_{j-1} \cdots V_i h_i V_i^\dagger \cdots V_{j-1}^\dagger, h_j\right] V_j^\dagger \cdots V_N^\dagger.
\end{align}
As in the main text, we are lead to compute $\mathbbm{E} \, \|(\mathbbm{1}_m - \Pi_n) K_N \Pi_n x \|^2$. When averaging over the noise $\Phi$ and taking into account the independence of the matrices $h_i$, we obtain:
\begin{align}
\mathbbm{E}_\Phi \, \|(\mathbbm{1}_m - \Pi_n) K_N \Pi_n x \|^2 &= \mathbbm{E}_\Phi \, \|(\mathbbm{1}_m - \Pi_n) H_N \Pi_n x \|^2 \nonumber\\
& \, + \frac{\epsilon^2}{4} \mathbbm{E}_\Phi \,  \sum_{1 \leq i < j \leq N} \| (\mathbbm{1}_m - \Pi_n)  V_N \cdots V_j \left[ V_{j-1} \cdots V_i h_i V_i^\dagger \cdots V_{j-1}^\dagger, h_j\right] V_j^\dagger  \cdots V_N^\dagger \Pi_n x \|^2
\end{align}
where we used that $\mathbbm{E}_\Phi \, h_i = \mathbbm{E}_\Phi \, h_i h_j h_k = 0$.
First, we notice that 
\begin{align}
\mathbbm{E}_\Phi \, \|(\mathbbm{1}_m - \Pi_n) K_N \Pi_n x \|^2 \geq \mathbbm{E}_\Phi \, \|(\mathbbm{1}_m - \Pi_n) H_N \Pi_n x \|^2,
\end{align}
meaning that the bound derived in the main text gives a correct upper bound on the permanent. 
Let us introduce the quantity
\begin{equation}
\mathcal{M}_{i,j} := 
\left\{
\begin{array}{ccc}
 \| (\mathbbm{1}_m - \Pi_n)  V_N \cdots V_j V_{j-1} \cdots V_i h_i V_i^\dagger \cdots V_{j-1}^\dagger h_j V_j^\dagger  \cdots V_N^\dagger \Pi_n x \|^2 & \:\text{if} \: &i <j \\ 
  \| (\mathbbm{1}_m - \Pi_n)  V_N \cdots V_j h_j V_{j-1} \cdots V_i h_i V_i^\dagger \cdots V_{j-1}^\dagger V_j^\dagger  \cdots V_N^\dagger \Pi_n x \|^2 & \: \text{if} \:& i >j \\
  \end{array}
  \right.
\end{equation}
so that
\begin{align}
\mathbbm{E}_\Phi \, \|(\mathbbm{1}_m - \Pi_n) K_N \Pi_n x \|^2 - \mathbbm{E}_\Phi \, \|(\mathbbm{1}_m - \Pi_n) H_N \Pi_n x \|^2 \leq \frac{\epsilon^2}{4} \sum_{ i\ne j} \mathcal{M}_{i,j}.
\end{align}
Consider the case $i < j$. Recalling that 
\begin{align}
\mathbbm{E}\Phi \, h_i^{\otimes 2}
& =: \sum_{l_1,l_2=1}^2  |i_{l_1} \rangle \langle i_{l_2}| \otimes |i_{l_2} \rangle \langle i_{l_1}|,
\end{align}
where $|i_1\rangle, |i_2\rangle$ refer respectively to the bottom left and top left input of the beamsplitter $V_i$, and using $\mathbbm{1}_m - \Pi_n \leq \mathbbm{1}_m$, we obtain:
\begin{align}
\mathcal{M}_{i,j} &\leq  \|   V_N \cdots V_j V_{j-1} \cdots V_i h_i V_i^\dagger \cdots V_{j-1}^\dagger h_j V_j^\dagger  \cdots V_N^\dagger \Pi_n x \|^2\\
&=  \|   h_i V_i^\dagger \cdots V_{j-1}^\dagger h_j V_j^\dagger  \cdots V_N^\dagger \Pi_n x \|^2\\
&= \langle x | \Pi_n V_N \cdots V_j h_j V_{j-1} \cdots V_i h_i^2  V_i^\dagger \cdots V_{j-1}^\dagger h_j V_j^\dagger  \cdots V_N^\dagger \Pi_n |x \rangle
\end{align}
Averaging over the choice of unitary $U^\mathrm{BS}$,
\begin{align}
\mathbbm{E}_{\mathrm{BS}} \,\mathcal{M}_{i,j}  &= \mathbbm{E}_{\mathrm{BS}} \mathbbm{E}_\Phi \mathbbm{E}_x \,\sum_{i_k, j_l} \langle x | \Pi_n V_N \cdots V_j |j_{l_1} \rangle \langle j_{l_2}| V_{j-1} \cdots V_i |i_{k_1} \rangle \langle i_{k_2}| V_i^\dagger \cdots V_{j-1}^\dagger |j_{l_2} \rangle \langle j_{l_1}| V_j^\dagger  \cdots V_N^\dagger \Pi_n |x \rangle\\
&= \frac{1}{n} \mathbbm{E}_{\mathrm{BS}}  \,\sum_{i_k, j_l}\langle j_{l_1}| V_j^\dagger  \cdots V_N^\dagger \Pi_n V_N \cdots V_j |j_{l_1} \rangle \langle j_{l_2}| V_{j-1} \cdots V_i |i_{k_1} \rangle \langle i_{k_2}| V_i^\dagger \cdots V_{j-1}^\dagger |j_{l_2} \rangle \\
&= \frac{1}{n} \mathbbm{E}_{\mathrm{BS}}  \,\sum_{k_1,l_1,l_2=1}^2 \langle j_{l_1}| V_j^\dagger  \cdots V_N^\dagger \Pi_n V_N \cdots V_j |j_{l_1} \rangle\,  | \langle j_{l_2}| V_{j-1} \cdots V_i |i_{k_1} \rangle |^2 
\end{align}
where we used in the last equality that $\mathbbm{E}_\mathrm{BS}\,  \langle i_{1}| V_i^\dagger \cdots V_{j-1}^\dagger |j_{l_2} \rangle \langle j_{l_2}| V_{j-1} \cdots V_i |i_{2} \rangle =0$.

Let us compute the different terms:
\begin{align}
 \mathbbm{E}_{\mathrm{BS}}  \,\sum_{l_1=1}^2\langle j_{l_1}| V_j^\dagger  \cdots V_N^\dagger \Pi_n V_N \cdots V_j |j_{l_1} \rangle &=  \mathbbm{E}_{\mathrm{BS}} \langle j_{1}| V_j^\dagger  \cdots V_N^\dagger \Pi_n V_N \cdots V_j |j_{1} \rangle + \langle j_{2}| V_j^\dagger  \cdots V_N^\dagger \Pi_n V_N \cdots V_j |j_{2} \rangle\\
 & = 2\frac{n}{\max(n, j_2) }= 2\sum_{\alpha=1}^n  \mathrm{P}_\mathrm{BS} \left[ j \rightarrow \alpha\right]
\end{align}
where $ \mathrm{P}_\mathrm{BS} \left[ j \rightarrow \alpha\right]$ denotes the probability that a photon entering the network at beamsplitter $j$ exits in output mode $\alpha$ and where we recall that $j_2$ refers to the top left input of the beamsplitter $V_j$, and
\begin{align}
 \mathbbm{E}_{\mathrm{BS}}  \,\sum_{l,k=1}^2 | \langle j_{l}| V_{j-1} \cdots V_i |i_{k} \rangle |^2 
 &=  \mathbbm{E}_{\mathrm{BS}}  \,\sum_{l,k=1}^2 | \langle j_{l}| V_{j-1} \cdots V_{i+1}  |i_{k} \rangle |^2 \\
 &= \mathbbm{P}_\mathrm{BS} \left[ i \rightarrow j \right]
\end{align}
where $ \mathbbm{P}_\mathrm{BS}\left[ i \rightarrow j \right]$ is the probability that a photon entering the network at beamsplitter $i$ exits the network at beamsplitter $j$.
As usual, let us denote by $(i_1, i_2)$ the ``coordinates'' of $i$, that is, the index of the first input mode and the second input mode (with $i_2 >i_1$) and by $(j_1, j_2)$ those of beamsplitter $j$. 

Putting these together, one obtains:
\begin{align}
\mathbbm{E} \, \sum_{i <j} \mathcal{M}_{i,j} &= \frac{2}{n} \sum_{i=1}^N \sum_{j=1}^{i-1} \sum_{\alpha \leq n}  \mathbbm{P}_\mathrm{BS}\left[ i \rightarrow j \right] \mathrm{P}_\mathrm{BS} \left[ j \rightarrow \alpha\right]\\
&= \frac{2}{n} \sum_{i=1}^N\sum_{\alpha \leq n}  \mathbbm{P}_\mathrm{BS}\left[ i \rightarrow \alpha\right] \times \left|\left\{\text{vertical lines beween $i$ and $\alpha$} \right\}\right|\\
&= \frac{2}{n} \sum_{i=1}^N\sum_{\alpha \leq n}  \mathbbm{P}_\mathrm{BS}\left[ i \rightarrow \alpha\right] (i_1 + i_2-\alpha-2)\\
&= \frac{2}{n} \sum_{i=1}^N\sum_{\alpha = 1}^{\min(n,i_2)}\frac{1}{i_2+1} (i_1 + i_2-\alpha-2)\\
&= \frac{2}{n} \sum_{i_2=1}^m \sum_{i_1=1}^{\min(n, i_2-1)} \sum_{\alpha = 1}^{\min(n,i_2)}\frac{1}{i_2+1} (i_1 + i_2-\alpha-2)\\
&= \frac{2}{n} \sum_{i_2=1}^n \sum_{i_1=1}^{i_2-1} \sum_{\alpha = 1}^{i_2}\frac{1}{i_2+1} (i_1 + i_2-\alpha-2)
+ \frac{2}{n} \sum_{i_2=n+1}^m \sum_{i_1=1}^{n} \sum_{\alpha = 1}^{n}\frac{1}{i_2+1} (i_1 + i_2-\alpha-2)\\
&= \frac{2}{n} \sum_{i_2=1}^n \sum_{i_1=1}^{i_2-1} \sum_{\alpha = 1}^{i_2}\frac{i_2-3}{i_2+1} 
+ \frac{2}{n} \sum_{i_2=n+1}^m \sum_{i_1=1}^{n} \sum_{\alpha = 1}^{n}\frac{i_2-2}{i_2+1}\\
&\leq 2 \sum_{i_2=1}^n \sum_{i_1=1}^{i_2} \sum_{\alpha = 1}^{i_2}1 
+ \frac{2}{n} \sum_{i_2=n+1}^m \sum_{i_1=1}^{n} \sum_{\alpha = 1}^{n}1\\
&\leq \frac{2}{n} \sum_{i_2=1}^ni_2^2 + 2(m-n) n^2\\
&\leq 2 m n
\end{align}

The sum of the $\mathcal{M}_{i,j}$ for $i>j$ can be handled exactly in the same way. 
Finally, we obtain the bound: 
\begin{align}
\mathbbm{E}_\Phi \, \|(\mathbbm{1}_m - \Pi_n) K_N \Pi_n x \|^2 - \mathbbm{E}_\Phi \, \|(\mathbbm{1}_m - \Pi_n) H_N \Pi_n x \|^2 \leq mn \epsilon^2.
\end{align}
Recalling that $\mathbbm{E}_\Phi \, \|(\mathbbm{1}_m - \Pi_n) H_N \Pi_n x \|^2 = O( n \ln n)$, we conclude that the higher order terms of the Baker-Campbell-Haussdorf formula can indeed be neglected in the regime where $n\epsilon \ll 1$.

\section{Study of a random network of beamsplitters}
\label{reck}

In the \textsc{BosonSampling} problem, the unitary $U^\mathrm{BS}$ is sampled from the Haar measure on the unitary group $U(m)$. 
We wish to understand the induced distribution on $U(2)$ for each of the $\tbinom{m}{2}$ beamsplitters in the implementation of $U^\mathrm{BS}$ obtained by the procedure of Reck et al \cite{RZB94}.

\begin{figure}
\centering
 \includegraphics[width=100mm]{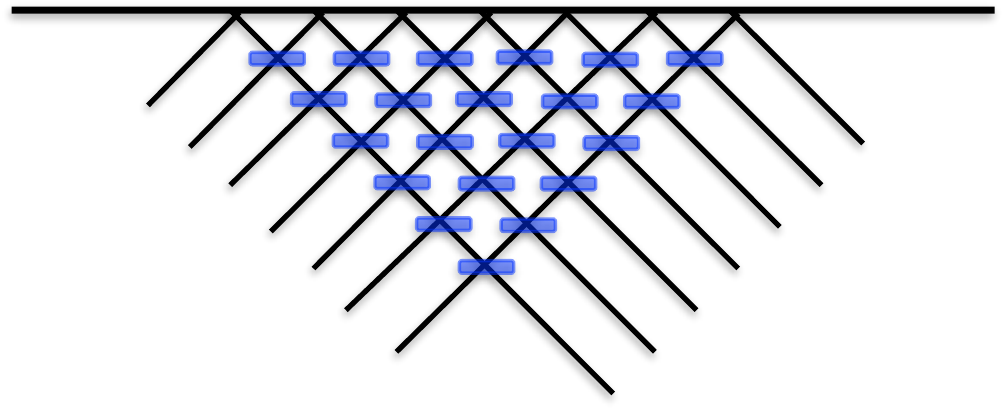}
  \fcaption{When averaging over the Haar measure over the unitary group, the expected transmissivity $T$ and reflectivity $R$ of a given beamsplitter only depend on their its in the network: the beamsplitters of the first line (from the top) have $T=R=1/2$, those of the $k^\mathrm{th}$ line have $T_k = \frac{k}{k+1}$ and $R_k = \frac{1}{k+1}$. With these notations, the height of beamsplitter $\{k_1,k_2\}$ is $k_2-k_1$. In other words, $T_{k_1, k_2} = \frac{k_2-k_1}{k_2-k_1+1}$ and $R_{k_1,k_2} = \frac{1}{k_2-k_1+1}$.
  }
   \label{figure:simple-network}
\end{figure}

Let us first consider the transmissivity $T_{k_1,k_2}$ and reflectivity $R_{k_1,k_2}$ for the beamsplitter acting on modes $k_1$ and $k_2 >k_1$, when averaging over the choice of unitary $U^\mathrm{BS}$. We recall that input modes are labelled $1$ to $m$ starting at the bottom (left) of Fig.~\ref{figure:simple-network}. 
It is easy to compute these quantities by considering the input $|1,0, \ldots, 0\rangle$ corresponding to a photon in the first mode and vacuum in the remaining modes. 
When averaging over the choice of unitary $U^\mathrm{BS}$, the probability that the photon is detected in a given output mode should be $1/m$. This leads immediately to 
\begin{align}
T_{1,k_2} = \frac{k_2-1}{k_2}, \quad R_{1, k_2} = \frac{1}{k_2}.
\end{align}
One can then proceed by induction and consider the input state $|0, 1, 0, \ldots, 0\rangle$. Again, when averaging over the choice of unitary, the probability of detecting the photon in any of the output modes should be $1/m$, which leads to $R_{2,k_2} = 1-T_{2,k_2} = \frac{1}{k_2-1}$. 
The general case is obtained similarly:  $R_{k_1,k_2} = 1-T_{k_1,k_2} = \frac{1}{k_2-k_1+1}$.

\section{Computation of $p_1^k =  \langle e_{k_1} |P_n^k | e_{k_1} \rangle$}
\label{computation_p_1}

\begin{figure}
\centering
 \includegraphics[width=100mm]{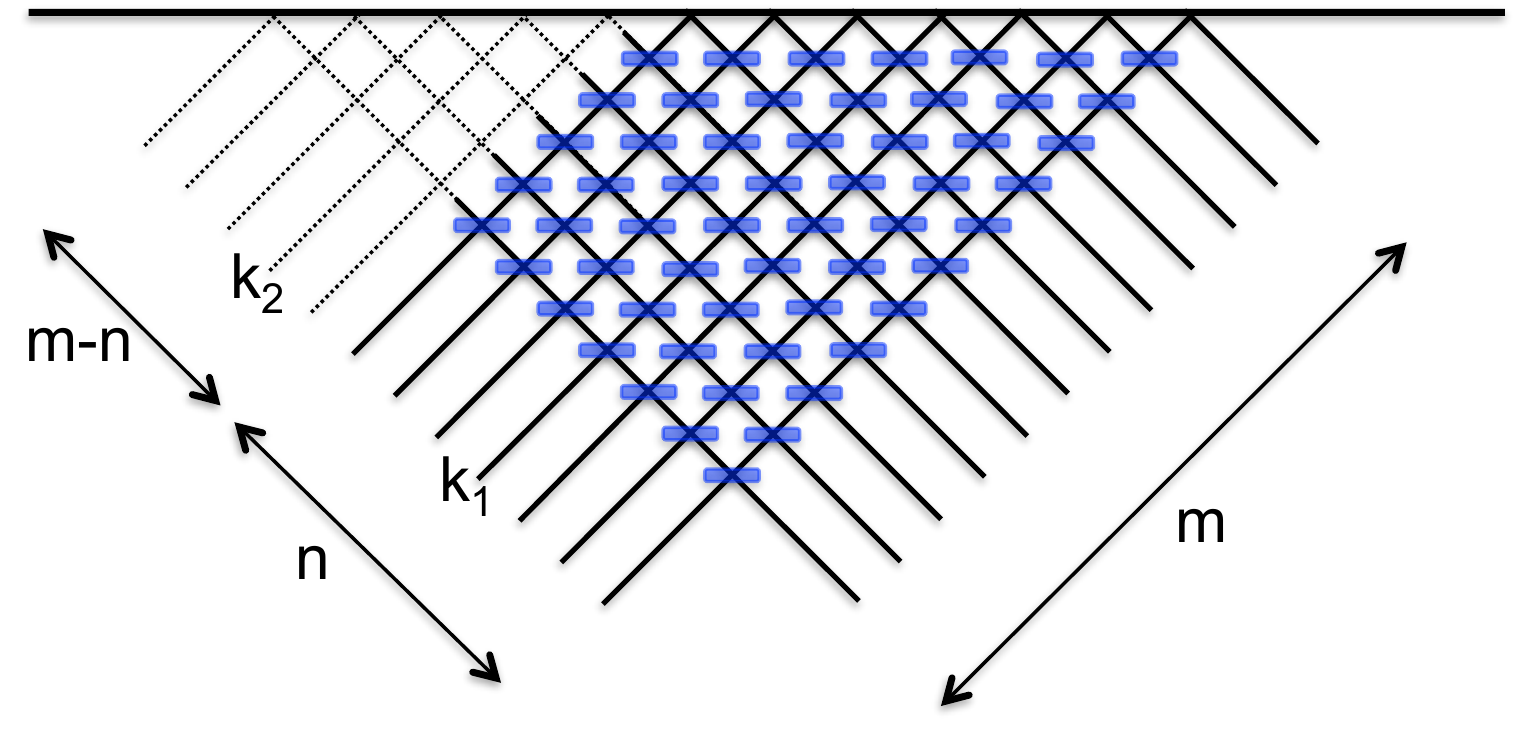}
 \includegraphics[width=100mm]{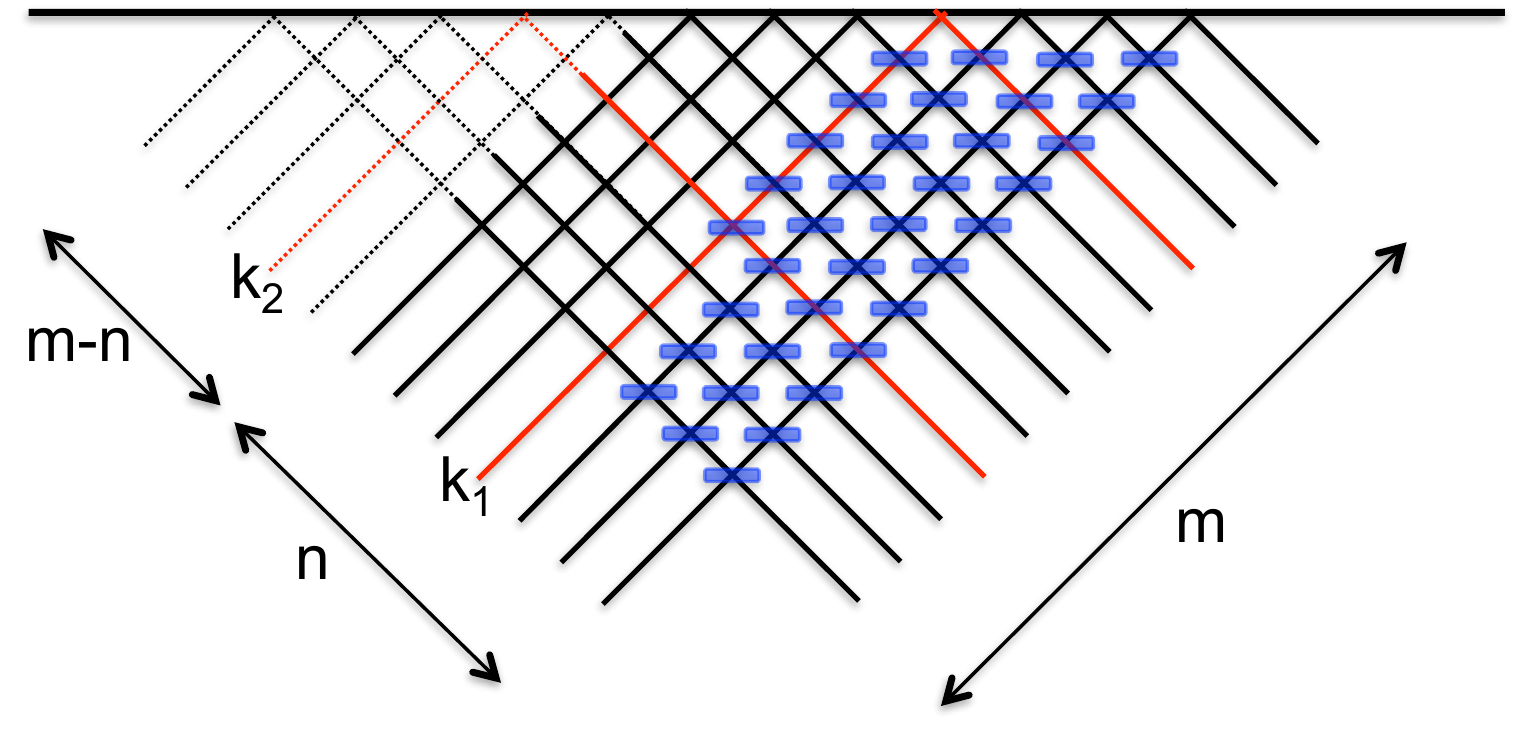}
  \includegraphics[width=100mm]{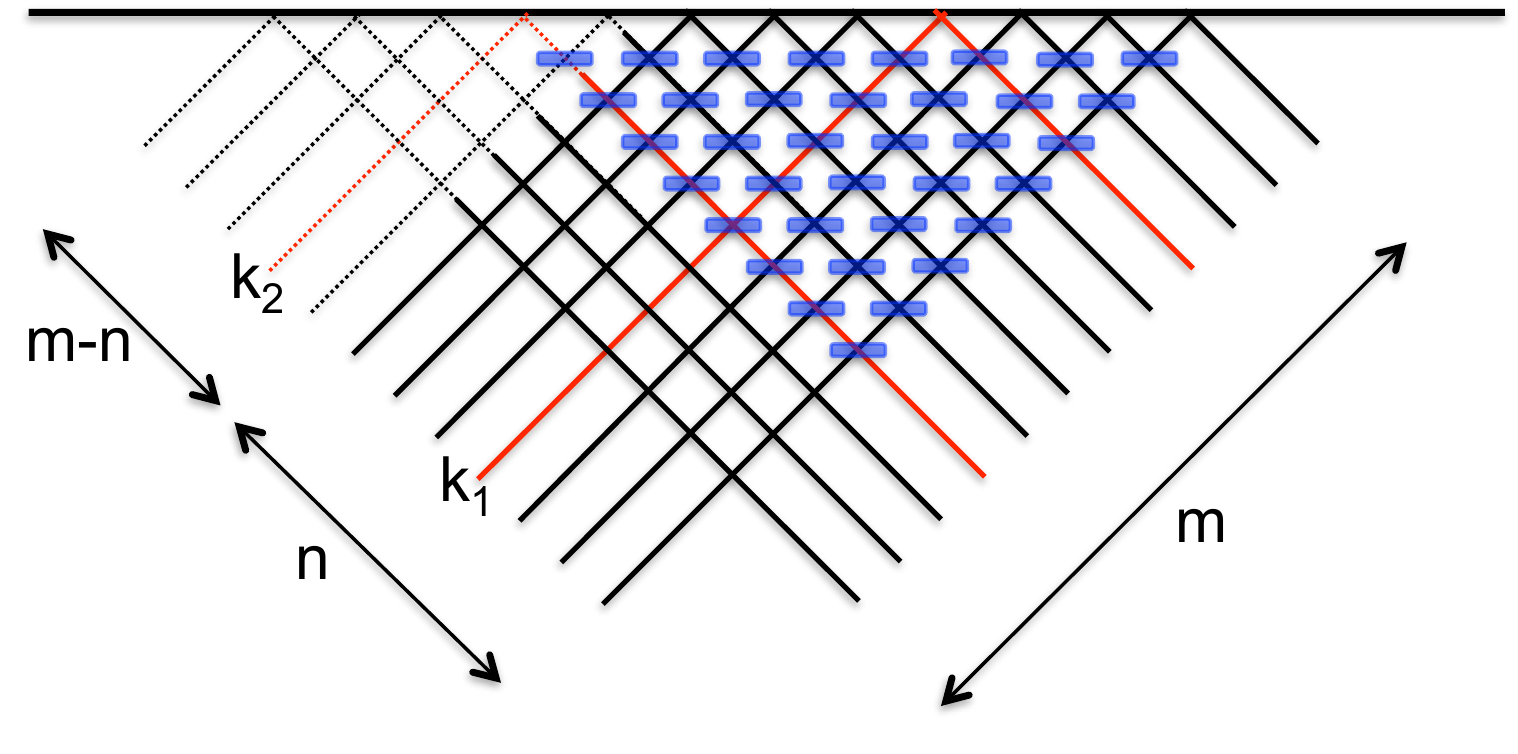}
    \fcaption{On the top figure, we represent the network of beamsplitters corresponding to $V_1$ up to $V_N$, as well as the specific modes $k_1$ and $k_2$. In order to compute $ \langle e_{k_1} |P_n^k | e_{k_1} \rangle $, one should first only consider beamsplitters of index larger than $k$ (middle figure). Then, if the input is a single photon in mode $k_1$, the networks depicted in the middle and bottom figures are equivalent. Moreover, if the initial network of beamsplitters is drawn from the Haar measure on $U(m)$ then it action on a state containing one single photon in mode $k_1$ is the same as the action of the network of the bottom figure, if it is drawn from the Haar measure on $U(k_2)$.}
   \label{figure:network-proba}
\end{figure}

Let us consider the index $k$ such that $V_k$ acts on modes $k_1 \in [1, n]$ and $k_2 \in [n+1, m]$. 
The quantity $p_1^k$ is the probability of detecting a photon in one of the first $n$ output modes, when inputting a single photon in mode $k_1$ of a network of beamsplitters implementing $V_{k}, V_{k+1}, \ldots, V_N$ and can be written:
\begin{align}
 \langle e_{k_1} |P_n^k | e_{k_1} \rangle = \mathrm{tr} \left[\Pi_n V_N \cdots V_k |e_{k_1}\rangle \langle e_{k_1} | V_k^\dagger \cdots V_N^\dagger \right],
 \end{align}
 where $V_k$ up to $V_N$ are represented in blue in the middle figure of Fig.~\ref{figure:network-proba}.

Given that beamsplitters acting only on vacuum modes can be replaced by the identity, it is straightforward to see that if the input of the network is a single photon in mode $k_1$, then the networks depicted in the bottom two figures of Fig.~\ref{figure:network-proba} are equivalent. 
In particular, $\mathbbm{E}_\mathrm{BS} V_N \cdots V_k |e_{k_1}\rangle \langle e_{k_1}| V_k^\dagger \cdots V_N^\dagger = \frac{1}{k_2} \Pi_{k_2}$, which gives 
\begin{align}
\mathbbm{E}_\mathrm{BS}\, p_1^k = \frac{n}{\max(n,k_2)}.
\end{align}
By symmetry, one also has $\mathbbm{E}_\mathrm{BS}\, p_2^k = \frac{n}{\max(n,k_2)}$.

In order to compute the second moment of $p_i^k$, we proceed differently. For $U^\mathrm{BS}$ acting on $m$ modes, let us write $\tilde{U}$ for the unitary acting on $k_2$ modes corresponding to the array of beamsplitters depicted on the bottom figure of Fig.~\ref{figure:network-proba} (that is, when removing all the beamsplitters of index $<k$).
The probability $p_1^k$ is given by:
\begin{align}
p_1^k = \sum_{i=1}^n \left| \tilde{U}_{k_1,i} \right|^2.
\end{align}
Since $\tilde{U}$ is drawn from the Haar measure on $U(k_2)$, the vector $(x_1, \ldots, x_{k_2})$ defined by $x_i =  \left|\tilde{U}_{k_1,i} \right|^2$ is uniformly distributed on the unit sphere of $\mathbbm{C}^{k_2}$, and $p_1^k = \sum_{i=1}^n |x_i|^2$. 
The vector $x$ can be obtained from $k_2$ independent Gaussian random variables $y_1, \ldots, y_{k_1} \sim \mathcal{N}(0, 1)_{\mathbbm{C}}$, via $x_i = y_i / \sqrt{\sum_{i=1}^{k_2} |y_i|^2}$, from which one infers that:
\begin{align}
p_1^k =  \frac{\sum_{i=1}^n |y_i|^2}{\sum_{i=1}^{k_2} |y_i|^2}.
\end{align} 
Computing the second moment of this random variable, we obtain:
\begin{align}
\mathbbm{E}_\mathrm{BS}\,  \left(p_1^k\right)^2 &= \frac{n(n+2)}{k_2(k_2+2)}.
\end{align}
By symmetry, $\mathbbm{E}_\mathrm{BS}\, \left(p_2^k\right)^2 = \mathbbm{E}_\mathrm{BS}\,  \left(p_1^k\right)^2$.

\section{Chernoff and Gaussian Matrix series}
\label{chernoff}

In this section, we compute a bound valid with large probability on the largest singular value of $(1- \Pi_n) H_N\Pi_n$. This will allow us to derive the following inequality:
\begin{align}
\mathbbm{P}_{\Phi, \mathrm{BS}} \left[ \|(\mathbbm{1}_m -\Pi_n) H_N \Pi_N x \|^2 \geq  12 n \ln \frac{em}{n} \ln \frac{m}{\delta} \right] \leq \delta.
\end{align}

For this, we study the largest singular value of $H_N = \sum g_k$, which gives an upper bound for $\|(\mathbbm{1}_m -\Pi_n) H_N \Pi_N x \|$. There are two sources of randomness to be considered: the randomness linked to the noise $\Phi$ is dealt with a Gaussian matrix series while the randomness of the \textsc{BosonSampling} unitary is dealt with a matrix Chernoff bound. 
In both cases, we will use matrix concentration inequalities due to Tropp \cite{Tro12}.

Let us first take care of the randomness due to the noise $\Phi$.
Consider the sum:
\begin{align}
H_N = \sum_{k=1}^N V_N \cdots V_k h_k V_k^\dagger \cdots V_N^\dagger
\end{align}
where the unitaries $\{V_i\}$ are constants. 

Let us recall Theorem 4.1.1 of Ref.~\cite{Tro12}.
\begin{theorem}{Matrix Gaussian Series.}
\label{tropp-Gaussian}
Consider a finite sequence $\{A_k\}$ of fixed Hermitian matrices with dimension $m$, and let $\{\gamma_k\}$ be a finite sequence of independent normal variables. Form the matrix Gaussian series 
$$Y = \sum_k \gamma_k A_k. $$ 
Then, for all $t\geq 0$, 
\begin{align}
\mathbbm{P} \left[\lambda_{\mathrm{max}}(Y)\geq t \right] \leq m \exp(-t^2/2\sigma^2),
\end{align}
where $\sigma^2 = \left\|\mathbbm{E} (Y^2) \right\|$, where $\|\cdot\|$ refers to the spectral norm.
\end{theorem}

Here, in order to get a bound for the largest eigenvalue of $H_N$, we need to compute the variance parameter:
\begin{align}
 \sigma^2 &:=  \|\mathbbm{E}_{\Phi}  H_N^2 \| \\
 &= 2\left\| \sum_{k=1}^N V_N \cdots V_k |k+\rangle \langle k+| V_k^\dagger \cdots V_N^\dagger\right\|
 \end{align}
 where we recall that $\mathbbm{E}_\Phi h_k^2 = (|e_{k_1}\rangle + |e_{k_2}\rangle)(\langle e_{k_1}| +\langle e_{k_2}|)
 = 2 |k+ \rangle \langle k+|$ where we define $|k+\rangle := \frac{1}{\sqrt{2}}(|e_{k_1}\rangle + |e_{k_2}\rangle)$.

We will use a matrix Chernoff bound in order to derive an upper bound on $\sigma^2$. Indeed, $\sigma^2$ can be seen as the spectral norm of a sum of random Hermitian matrices given by $X_k =2V_N \cdots V_k |k+\rangle \langle k+| V_k^\dagger \cdots V_N^\dagger$.
Let us recall another result of Tropp \cite{Tro12}:
\begin{theorem}
\label{tropp-chernoff}
Consider a finite sequence $\{X_k\}$ of independent, random, Hermitian matrices of dimension $m$ that satisfy $X_k \succeq 0$ and $\lambda_\mathrm{max} (X_k) \leq 2$. 
Define the random matrix $Y = \sum_k X_k$. Then, for $\delta \geq 0$,
\begin{align}
\mathbbm{P}\left[ \lambda_\mathrm{max} (Y) \geq (1+\delta) \mu_\mathrm{max} \right] \leq m \left[\frac{e^\delta}{(1+\delta)^{1+\delta}} \right]^{\mu_\mathrm{max}/2},
\end{align}
where $\mu_\mathrm{max} = \lambda_\mathrm{max}(\mathbbm{E}Y)$.
\end{theorem}

Let us therefore introduce $\mu_{\mathrm{max}} = \lambda_{\mathrm{max}} (\mathbbm{E}_{\mathrm{BS}} \sum_{k=1}^N X_k)$:
\begin{align}
\mathbbm{E}_{\mathrm{BS}} \sum_{k=1}^N X_k &= \mathbbm{E}_{\mathrm{BS}} \sum_{k=1}^N V_N \cdots V_k |e_{k_1}\rangle \langle e_{k_1}| V_k^\dagger \cdots V_N^\dagger + \sum_{k=1}^N V_N \cdots V_k |e_{k_2}\rangle \langle e_{k_2}| V_k^\dagger \cdots V_N^\dagger \\
&= \sum_{k_1=1}^n \sum_{k_2=k_1+1}^{m} \frac{2}{k_2} \Pi_{k_2} \\
&=2 \sum_{k_1=1}^n \sum_{k_2=k_1+1}^{m} \frac{1}{k_2} \sum_{l=1}^{k_2} |e_l\rangle \langle e_l|\\
&=2 \sum_{k_2=1}^m  \frac{1}{k_2} \sum_{l=1}^{k_2}  \sum_{k_1=1}^{\min(k_2,n)}|e_l\rangle \langle e_l|\\
&=2 \sum_{k_2=1}^m  \frac{\min(k_2,n)}{k_2} \sum_{l=1}^{k_2}  |e_l\rangle \langle e_l|\\
&=2 \sum_{k_2=1}^n   \sum_{l=1}^{k_2}  |e_l\rangle \langle e_l| + 2n \sum_{k_2=n+1}^m  \frac{1}{k_2} \sum_{l=1}^{k_2}  |e_l\rangle \langle e_l|
\end{align}
It is easy to compute the maximum eigenvalue of this matrix:
\begin{align}
\mu_\mathrm{max} &= 2n + 2n \sum_{i=n+1}^m \frac{1}{i}
\end{align}
which leads to 
\begin{align}
2n\left(1 + \ln \frac{m}{n+1} +\frac{1}{2m+1} - \frac{1}{2n+2} \right) \leq \mu_\mathrm{max} \leq 2n\left(1+\ln \frac{m}{n+1} +\frac{1}{2m} - \frac{1}{2(n+1)}\right)
\end{align}
and
\begin{align}
2n \ln \frac{m}{n+1}  \leq \mu_\mathrm{max} \leq 2n \ln \frac{em}{n}.
\end{align}

Take $\delta = 2$ so that $e^\delta/(1+\delta)^{1+\delta} \leq 1/e$, and apply the matrix Chernoff bound of Theorem \ref{tropp-chernoff}:
\begin{align}
\mathbbm{P}_{\mathrm{BS}} \left[\sigma^2 \geq6n \ln\frac{em}{n} \right] \leq m \left(\frac{n+1}{m} \right)^{n}.
\end{align}

For all practical purposes, i.e. reasonable values of the parameters $n$ and $m$, this probability is negligible, and in the following, we can assume that $6n \ln (em/n)$ is a hard upper bound on $\sigma^2$.

We are now ready to apply Theorem \ref{tropp-Gaussian} with $\sigma^2 = 6n \ln \frac{em}{n}$: for any $\delta \geq 0$,
\begin{align}
\mathbbm{P}_\Phi \left[ \lambda_\mathrm{max} (H_N) \geq  \sqrt{2 \sigma^2 \ln \frac{m}{\delta}} \right] \leq \delta,
\end{align}
which leads to 
\begin{align}
\mathbbm{P}_\Phi \left[ \lambda_\mathrm{max} (H_N) \geq  \sqrt{12 n \ln \frac{em}{n} \ln \frac{m}{\delta}} \right] \leq \delta.
\end{align}
In turn, for any unit vector $x$, one gets
\begin{align}
\mathbbm{P}_{\Phi, \mathrm{BS}} \left[ \|(\mathbbm{1}_m -\Pi_n) H_N \Pi_N x \|^2 \geq  12 n \ln \frac{em}{n} \ln \frac{m}{\delta} \right] \leq \delta.
\end{align}

\end{document}